\documentclass[12pt,a4paper]{article}

\pdfoutput=1
\usepackage{jheppub}
\usepackage{graphicx} % Required for inserting images
\usepackage[compat=1.1.0]{tikz-feynhand}
\usepackage{bm}
\usepackage{ascmac}
\usepackage{subcaption}
\usepackage{physics}
\usepackage[normalem]{ulem}
\usepackage{hyperref}

\newcommand{\MSbar}{\overline{\mathrm{MS}}}

\makeatletter
\gdef\@fpheader{}
\makeatother

\preprint{KEK--TH--2695, OU-HET-1268}

\title{Electroweak baryogenesis in 2HDM \\ without EDM cancellation}

\author[a]{Masashi Aiko}
\author[b,c,d]{Motoi Endo}
\author[e]{Shinya Kanemura}
\author[b,e]{Yushi Mura}

\affiliation[a]{National Institute of Technology, Miyakonojo College, Miyakonojo, Miyazaki 885-8567, Japan}
\affiliation[b]{KEK Theory Center, Tsukuba, Ibaraki 305--0801, Japan}
\affiliation[c]{Graduate Institute for Advanced Studies, SOKENDAI, Tsukuba, Ibaraki 305--0801, Japan}
\affiliation[d]{Kobayashi-Maskawa Institute (KMI) for the Origin of Particles and the Universe, Nagoya University, Nagoya 464--8602, Japan}
\affiliation[e]{Department of Physics, Osaka University, Toyonaka, Osaka 560-0043, Japan}

\begin{document}

\abstract{
We study two Higgs doublet models with successful electroweak baryogenesis but without cancellations of electric dipole moments (EDMs).
For the baryogenesis, additional scalar bosons are favored to couple mainly with the top quark with CP violations.
However, if they also couple to light fermions of the Standard Model, the model is limited severely by EDMs, and additional CP phases irrelevant to the baryogenesis are often introduced to cancel the contributions to the EDMs.
Alternatively, we consider a scenario where the light-fermion couplings are suppressed to avoid the constraints. 
In our scenario, it is found that the leading contributions arise in the top-quark EDMs at the two-loop level.
They induce the electron, neutron, and proton EDMs via radiative corrections.
Since there is no additional CP-violating phase, they are correlated with the baryon asymmetry.
We show that our scenario is compatible with the current experimental bounds and is within the scope of future EDM experiments.
}

\maketitle
\section{Introduction \label{sec:intro}}

The Standard Model (SM) is a successful theory of particle physics.
However, there are unsolved problems, such as the baryon asymmetry of the Universe (BAU).
Since the baryon asymmetry is smeared during the inflation era, baryogenesis should occur during the thermal history of the Universe.

Electroweak baryogenesis (EWBG)~\cite{Kuzmin:1985mm} is a promising scenario of baryogenesis, which is well motivated by its testability.
In the scenario, the BAU is generated by electroweak (EW) physics.
Among the Sakharov conditions~\cite{Sakharov:1967dj}, the baryon number is violated at finite temperature via the sphaleron process~\cite{Manton:1983nd}.
Also, C and CP violations can be realized by EW chiral interactions with CP phases.
The out-of-thermal equilibrium condition is satisfied around fast-moving bubble walls created by first-order EW phase transition.

Within the SM, however, EWBG cannot be realized. 
The CP violation in the quark sector~\cite{Cabibbo:1963yz,Kobayashi:1973fv} is too small to explain the observed BAU~\cite{Gavela:1993ts,Huet:1994jb,Gavela:1994ds,Gavela:1994dt}.
In addition, the EW phase transition in the SM is not first order but crossover~\cite{Kajantie:1996mn,DOnofrio:2015gop}.
Therefore, a model for EWBG should include new CP violations and mechanisms making the EW phase transition first order.

Despite many experimental efforts, detailed properties of the Higgs sector are still unknown. 
Hence, the Higgs sector may involve rich structures.
The two Higgs doublet model (2HDM) is one of the most straightforward extensions of the Higgs sector, in which EWBG has been studied for a long time~\cite{Turok:1990zg,Cline:1995dg,Cline:2011mm,Tulin:2011wi,Liu:2011jh,Ahmadvand:2013sna,Chiang:2016vgf,Guo:2016ixx,Fromme:2006cm,Dorsch:2016nrg,Basler:2021kgq,Aoki:2023xnn,Fuyuto:2017ewj,Modak:2018csw,Fuyuto:2019svr,Enomoto:2021dkl,Enomoto:2022rrl,Kanemura:2023juv,Goncalves:2023svb,Athron:2025iew}.
The first-order phase transition is caused by non-decoupling effects of additional scalar bosons~\cite{Funakubo:1993jg,Davies:1994id,Cline:1996mga}.
Their effects could be probed e.g.~by measuring the Higgs triple coupling~\cite{Kanemura:2004ch,Kanemura:2002vm,Kanemura:2004mg,Braathen:2019pxr,Braathen:2019zoh,Bahl:2022jnx}, the Higgs di-photon decay~\cite{Ellis:1975ap,Shifman:1979eb, Gavela:1981ri, Barroso:1999bf, Arhrib:2003vip, Djouadi:2005gj, Akeroyd:2007yh, Posch:2010hx, Kanemura:2016sos, Degrassi:2023eii, Aiko:2023nqj}, and the stochastic gravitational waves~\cite{Grojean:2006bp,Espinosa:2010hh,Caprini:2015zlo, Kakizaki:2015wua, Hashino:2016rvx}.

Also, new CP phases may contribute to CP-violating (CPV) observables such as electric dipole moments (EDMs)~\cite{Ecker:1983dj,Barr:1990vd}.
Currently, there are upper bounds on the EDMs for several elementary or composite particles (e.g.~electron~\cite{ACME:2018yjb,Roussy:2022cmp}, neutron~\cite{Abel:2020pzs}, and proton~\cite{Graner:2016ses,Sahoo:2016zvr}).
For example, JILA reported an upper bound on the electron EDM, $|d_e| \le 4.1 \times 10^{-30}~e\,\mathrm{cm}$ at 90\% confidence level~\cite{Roussy:2022cmp}.

The EDM has given strong constraints on scenarios of EWBG in the 2HDM.
In the softly broken $\mathbb{Z}_2$ symmetric 2HDM, where flavor-changing neutral current (FCNC) processes are naturally suppressed, it is challenging to produce the BAU sufficiently~\cite{Fromme:2006cm,Dorsch:2016nrg,Basler:2021kgq,Aoki:2023xnn}
\footnote{
This conclusion is given in the semi-classical approach~\cite{Cline:2000nw,Fromme:2006wx,Joyce:1994fu,Joyce:1994zn,Joyce:1994zt,Cline:2020jre}.
Although the VEV insertion approximation~\cite{Riotto:1995hh,Riotto:1997vy} may predict larger BAU~\cite{Basler:2021kgq,Cline:2021dkf}, theoretical issues have recently been pointed out for this approach~\cite{Kainulainen:2021oqs,Postma:2022dbr}.}.
When the $\mathbb{Z}_2$ symmetry is exact, there are no CP phases for the BAU, while a single CP phase can be introduced by softly breaking the symmetry.
Although the phase could contribute to the BAU, it induces the EDMs inevitably via Barr--Zee type diagrams~\cite{Barr:1990vd}, exceeding the experimental bounds.

The EDM constraints may be ameliorated if the $\mathbb{Z}_2$ symmetry is, {\it a priori}, not imposed.
Such a model has been studied to explain the BAU~\cite{Fuyuto:2019svr,Enomoto:2021dkl,Enomoto:2022rrl}.
One can introduce multiple CP phases in the Yukawa interactions and the scalar potential. 
Some could generate the BAU and EDMs simultaneously, and the others may also induce the EDMs, even if they are not relevant to the BAU. 
Hence, the Barr--Zee contributions to the EDMs could be canceled out by tuning the CP phases irrelevant to the BAU~\cite{Bian:2014zka, Cheung:2020ugr, Kanemura:2020ibp}.

In this paper, we would like to stress that such cancellations are not the unique approach to avoid the EDM bounds. 
The Barr--Zee contributions to the electron and light-quark EDMs depend on couplings of the additional scalar bosons with the SM light fermions. 
Hence, if we turn off those interactions, the EDM contributions are suppressed automatically. 
On the other hand, the BAU is mainly caused by the CPV interaction with the top quark~\cite{Zhang:1994fb,Fromme:2006wx}.
Therefore, the BAU can be produced sufficiently without being jeopardized by the EDMs if the light-fermion interactions are suppressed while keeping the top-quark coupling large. 
Such a setup cannot be accommodated to the softly broken $\mathbb{Z}_2$ symmetric 2HDM but could be realized in the general 2HDM.

We examine the contributions to the EDMs to check its viability. 
We focus on the CP phase of the top-quark interaction of the additional scalar bosons, which is relevant to the BAU, while the other CPV interactions irrelevant to the BAU are not introduced. 
Although the Barr--Zee contributions are absent by turning off the light-fermion couplings, it will be shown that the contributions can arise in the top-quark EDMs at the two-loop level.
They then induce the electron, neutron, and proton EDMs via radiative corrections.
Since there are no extra CP phases, the EDMs are correlated with the BAU.
We will show that our scenario is compatible with the current experimental bounds and is within the scope of future EDM experiments.

This paper is organized as follows.
In section~\ref{sec:model}, we introduce the general 2HDM and explain the renormalization scheme used in the effective potential.
In section~\ref{sec:EWBG}, we discuss which CP phases are essential for EWBG and specify a scenario for the analysis. 
The EDMs are studied in section~\ref{sec:EDM}.
We provide formulae of the top-quark EDMs, leading to the neutron, proton, and electron EDMs.
In section~\ref{sec:BAU_and_EDM}, numerical results of the BAU and the EDMs are shown.
We give discussion and conclusions in sections~\ref{sec:discussion} and \ref{sec:conclusion}, respectively.
In appendix~\ref{sec:Loopfunc}, definitions of the loop functions are given.
We explicitly show the renormalization scheme independence of the EDM formulae in appendix~\ref{sec:Renorm}.

\section{Two Higgs doublet model \label{sec:model}}

\subsection{Model \label{sec:lagrangian}}

Here, we discuss the general 2HDM. This model is composed of the two $\mathrm{SU}(2)_L$ doublet scalar fields $\Phi_k~(k=1,2)$ with the hypercharge $Y=1/2$.
The kinetic term of the scalar doublets is given by
\begin{align}
    \mathcal{L}_{\mathrm{kin}} = (D^\mu \Phi_1)^\dagger (D_\mu \Phi_1) + (D^\mu \Phi_2)^\dagger (D_\mu \Phi_2),
\end{align}
where the covariant derivative is defined by $D_\mu = \partial_\mu - ig^\prime \frac{1}{2} B_\mu - i g \frac{\sigma^a}{2} W_\mu^a$ with the Pauli matrices $\sigma^a$ ($a = 1,2,3$).

The Higgs potential is given by 
\begin{align}
    \mathcal{V} = &-\mu_1^2 (\Phi_1^\dagger \Phi_1) - \mu_2^2 (\Phi_2^\dagger \Phi_2) - \Big(\mu_3^2 (\Phi_1^\dagger \Phi_2) + \mathrm{h.c.} \Big) \notag \\
    &+\frac{1}{2}\lambda_1 (\Phi_1^\dagger \Phi_1)^2 +\frac{1}{2}\lambda_2 (\Phi_2^\dagger \Phi_2)^2 + \lambda_3 (\Phi_1^\dagger \Phi_1)(\Phi_2^\dagger \Phi_2) + \lambda_4 (\Phi_1^\dagger \Phi_2)(\Phi_2^\dagger \Phi_1) \notag \\
    &+ \bigg\{ \Big( \frac{1}{2}\lambda_5 \Phi_1^\dagger \Phi_2 + \lambda_6 \Phi_1^\dagger \Phi_1 + \lambda_7 \Phi_2^\dagger \Phi_2 \Big) \Phi_1^\dagger \Phi_2 + \mathrm{h.c.} \bigg\}.
    \label{eq:potential}
\end{align}
This potential is written in the Higgs basis~\cite{Davidson:2005cw}, where only $\Phi_1$ acquires the Vacuum Expectation Value (VEV).
We parametrize the two doublets as follows:
\begin{align}
    \Phi_1 = 
    \begin{pmatrix}
       G^+ \\
       \frac{1}{\sqrt{2}} ( v + h_1 + i G^0)
    \end{pmatrix},
    ~~
    \Phi_2 = 
    \begin{pmatrix}
       H^+ \\
       \frac{1}{\sqrt{2}} ( h_2 + i h_3)
    \end{pmatrix},
\end{align}
where $G^\pm$ and $G^0$ are the Nambu--Goldstone (NG) bosons, and $H^\pm$ and $h_{1,2,3}$ are the charged and neutral scalar bosons, respectively.
The parameters $\mu_1^2, \mu_2^2,\lambda_1,\lambda_2,\lambda_3,\lambda_4$ are real, and $\mu_3^2,\lambda_5,\lambda_6,\lambda_7$ are complex.
In the Higgs basis, we can redefine $\Phi_2$ as
\begin{align}
\Phi_2 \to e^{i \chi} \Phi_2
\label{eq:Higgs_basis_change}
\end{align}
with $\chi \in [0,2\pi)$. Under this transformation, the complex parameters are changed as 
\begin{align}
    \mu_3^2 \to e^{i \chi} \mu_3^2\qc
    \lambda_5 \to e^{2 i \chi} \lambda_5\qc
    \lambda_6 \to e^{i \chi} \lambda_6\qc
    \lambda_7 \to e^{i \chi} \lambda_7.
    \label{eq:rephasing1}
\end{align}
By using this degree of freedom, one of the complex phases can be absorbed into $\chi$.
The physical quantities are independent of the choice of $\chi$. 

The stationary conditions at the tree level are given by 
\begin{align}
    \eval{\frac{\partial \mathcal{V}}{\partial h_i}}_{\Phi_{k}=\expval{\Phi_{k}}} = 0 ~~ \Leftrightarrow ~~ \mu_1^2 = \frac{1}{2} \lambda_1 v^2,~ \mathrm{and}~ \mu_3^2 = \frac{1}{2} \lambda_6 v^2.
    \label{eq:stationary_condition}
\end{align}
The mass of the charged scalar bosons is given by 
\begin{align}
    m_{H^\pm}^2 = M^2 + \frac{1}{2} \lambda_3 v^2,
\end{align}
where we have defined $M^2 \equiv -\mu_2^2$.
The mass matrix of the neutral scalar bosons $h_{1,2,3}$ is obtained as
\begin{align}
    \mathcal{M}_{ij}^2
    &=
    \eval{\frac{\partial^2 \mathcal{V}}{\partial h_i \partial h_j}}_{\Phi_{k}=\expval{\Phi_{k}}}
    \notag \\
    &=
    \begin{pmatrix}
        \lambda_1 v^2 & \lambda_6^R v^2 & - \lambda_6^I v^2 \\
        \lambda_6^R v^2 & m_{H^\pm}^2 +\frac{1}{2} \big(\lambda_4 + \lambda_5^R \big) v^2 & -\frac{1}{2}\lambda_5^I v^2 \\
        - \lambda_6^I v^2 & -\frac{1}{2} \lambda_5^I v^2 & m_{H^\pm}^2 +\frac{1}{2}(\lambda_4 -  \lambda_5^R) v^2
    \end{pmatrix}.
\end{align}
The subscripts $i,j = 1,2,3$ are for the neutral scalar bosons, and $\star^R$ ($\star^I$) means the real (imaginary) part of the coupling constant.
The mass matrix depends on the basis, and we can take $\mathcal{M}_{23}^2 = \mathcal{M}_{32}^2 = 0$ with $\chi=-\mathrm{arg}[\lambda_5]/2$.
By rotating the basis of the neutral scalar bosons with an orthogonal matrix $\mathcal{R}$, we define the mass eigenbasis as
\begin{align}
    H_i = \mathcal{R}_{ij} h_j,
    \label{eq:Masseigen}
\end{align}
in which the mass matrix is diagonalized as 
\begin{align}
    \mathcal{R} \mathcal{M}^2 \mathcal{R}^T = \mathrm{diag}(m_{H_1}^2, m_{H_2}^2, m_{H_3}^2).
    \label{eq:relationRM}
\end{align}
We identify $H_1$ as the $125~\mathrm{GeV}$ Higgs boson, which was discovered at the LHC~\cite{ATLAS:2012yve,CMS:2012qbp}.
The orthogonal matrix $\mathcal{R}$ can be parametrized as~\cite{Haber:2006ue}
\begin{align}
    \mathcal{R}
    &=
    \mqty(
    c_{12} & -s_{12} & 0 \\
    s_{12} & c_{12} & 0 \\
    0 & 0 & 1)
    \mqty(
    c_{13} & 0 & -s_{13} \\
    0 & 1 & 0 \\
    s_{13} & 0 & c_{13})
    \mqty(
    1 & 0 & 0 \\
    0 & c_{23} & -s_{23} \\
    0 & s_{23} & c_{23})
    \notag \\
    &=
    \mqty(
    c_{13}c_{12} & -s_{12}c_{23}-c_{12}s_{13}s_{23} & -c_{12}s_{13}c_{23}+s_{12}s_{23} \\
    c_{13}s_{12} & c_{12}c_{23}-s_{12}s_{13}s_{23} & -s_{12}s_{13}c_{23}-c_{12}s_{23} \\
    s_{13} & c_{13}s_{23} & c_{13}c_{23}).    
    \label{eq:Rmatrix}
\end{align}
where $c_{ij}=\cos{\theta_{ij}}$ and $s_{ij}=\sin{\theta_{ij}}$.
The ranges of mixing angles are $-\pi \le \theta_{12}, \theta_{23} < \pi$, and $-\pi/2 \le \theta_{13} < \pi/2$.
The angle $\theta_{23}$ depends on the choice of $\chi$, while $\theta_{12}$ and $\theta_{13}$ do not~\cite{Haber:2006ue, Boto:2020wyf}.

The Yukawa interaction in the model is given by
\begin{align}
    \mathcal{L}_{Y} = - \sum_{k=1}^2 \sum_{l,m=1}^3 \left( \overline{Q^\prime_{l,L}} Y^{u}_{k,lm} \tilde{\Phi}_k u^\prime_{m,R} + \overline{Q^\prime_{l,L}} Y^{d}_{k,lm} \Phi_k d^\prime_{m,R} + \overline{L^\prime_{l,L}} Y^e_{k,lm} \Phi_k e^\prime_{m,R} + \mathrm{h.c.} \right),
\end{align}
where the subscripts $k=1,2$ and $l,m = 1,2,3$ are for the two scalar doublets and for the fermion flavors, respectively.
We here denote $\tilde{\Phi}_k = i\sigma_2 \Phi_k^*$.
The left-handed (right-handed) quark and lepton doublets, $Q_{l,L}^\prime$ ($u_{l,R}^\prime$ and $d_{l,R}^\prime$) and $L_{l,L}^\prime$ ($e_{l,R}^\prime$), are defined in the gauge eigenbasis.
By rotating in the flavor space, the Yukawa matrices for $\Phi_1$ can be diagonalized.
We define the unitary transformation in the flavor space as 
\begin{align}
    f_{l,L} = V^f_{L,lm} f_{m,L}^\prime, ~
    f_{l,R} = V^f_{R,lm} f_{m,R}^\prime,
\end{align}
where $f = u,d,e$, and the unitary matrices $V_{L,R}^f$ satisfy $V^f_L Y_1^f V_R^{f\dagger} = \mathrm{diag}(y_{f_1}, y_{f_2}, y_{f_3})$.
The Cabbibo--Kobayashi--Maskawa (CKM) matrix~\cite{Cabibbo:1963yz,Kobayashi:1973fv} is written by $V_{\mathrm{CKM}} = V_L^u V_L^{d \dagger}$.
On the other hand, the Yukawa matrices for $\Phi_2$ are generally not diagonalized by these unitary transformations.
We write the additional Yukawa matrices as $\rho^f = V_L^f Y_{2}^f V_R^{f \dagger}$, and their components are given by 
\begin{align}
    (\rho^u)_{lm} = 
    \begin{pmatrix}
    \rho_{uu} & \rho_{uc} & \rho_{ut} \\
    \rho_{cu} & \rho_{cc} & \rho_{ct} \\
    \rho_{tu} & \rho_{tc} & \rho_{tt}
    \end{pmatrix}, 
    ~~
    (\rho^d)_{lm} = 
    \begin{pmatrix}
    \rho_{dd} & \rho_{ds} & \rho_{db} \\
    \rho_{sd} & \rho_{ss} & \rho_{sb} \\
    \rho_{bd} & \rho_{bs} & \rho_{bb}
    \end{pmatrix},
    ~~
    (\rho^e)_{lm} = 
    \begin{pmatrix}
    \rho_{ee} & \rho_{e\mu} & \rho_{e\tau} \\
    \rho_{\mu e} & \rho_{\mu \mu} & \rho_{\mu \tau} \\
    \rho_{\tau e} & \rho_{\tau \mu} & \rho_{\tau \tau}
    \end{pmatrix}.
\end{align}
Under the basis transformation in eq.~\eqref{eq:Higgs_basis_change}, these matrices transform as 
\begin{align}
    \rho^u \to e^{- i \chi} \rho^u, ~~~~ \rho^D \to e^{i \chi} \rho^D,
    \label{eq:rephasing2}
\end{align}
where we have defined the superscript $D = d,e$ for the down-type fermions.
The off-diagonal elements of these matrices cause the FCNC processes, which are constrained by the flavor and collider experiments~\cite{Crivellin:2013wna}.

Finally, the CPV quantities, including the BAU and the EDM, are proportional to the imaginary part of the rephasing invariants.
Under the phase rotations in eq.~\eqref{eq:rephasing1}, there are three rephasing invariants:
\footnote{
We have omitted invariants including $\mu_3^2$ from the list because the phase of $\mu_3^2$ is the same as that of $\lambda_6$ with the stationary conditions in eq.~\eqref{eq:stationary_condition}.
}
\begin{align}
    \mathrm{Im}[\lambda_5^* \lambda_7^2], ~~ \mathrm{Im}[\lambda_5^* \lambda_6^2], ~~ \mathrm{Im}[\lambda_6 \lambda_7^*].
    \label{eq:rephasing_invariant1}
\end{align}
They are independent of the choice of $\chi$.
In the presence of the Yukawa matrices $\rho^f$, whose phase rotations are shown in eq.~\eqref{eq:rephasing2}, the additional rephasing invariants are obtained as
\begin{alignat}{3}
    &\mathrm{Im}[\lambda_5 (\rho^u)_{lm}^2], ~~& &\mathrm{Im}[\lambda_6 (\rho^u)_{lm}], ~~& &\mathrm{Im}[\lambda_7 (\rho^u)_{lm}], \notag \\
    &\mathrm{Im}[\lambda_5^* (\rho^{D})_{lm}^2], ~~& &\mathrm{Im}[\lambda_6^* (\rho^{D})_{lm}], ~~& &\mathrm{Im}[\lambda_7^* (\rho^{D})_{lm}], \notag \\
    &\mathrm{Im}[(\rho^u \rho^{D})_{lm}], ~~& &\mathrm{Im}[(\rho^u \rho^{u \dagger})_{lm}], ~~& &\mathrm{Im}[(\rho^D \rho^{D \dagger})_{lm}].
    \label{eq:rephasing_invariant2}
\end{alignat}
We note that the last two terms in the third line have a non-zero value only when $l \neq m$.

\subsection{Renormalization \label{sec:renormalization}}

We discuss the renormalization of the 2HDM based on the effective potential.
The relevant part of the Lagrangian is given by
\begin{align}
    \mathcal{L} &\supset
    |\partial_\mu \Phi_{1,B}|^2 + |\partial_\mu \Phi_{2,B}|^2 + \mu_{1,B}^2 (\Phi_{1,B}^\dagger \Phi_{1,B}) +  \mu_{2,B}^2 (\Phi_{2,B}^\dagger \Phi_{2,B}) \notag \\
    &+ \Big( \mu_{3,B}^2 (\Phi_{1,B}^\dagger \Phi_{2,B}) + \mathrm{h.c.} \Big) -\frac{1}{2} \lambda_{1,B} (\Phi_{1,B}^\dagger \Phi_{1,B})^2 -\frac{1}{2} \lambda_{2,B} (\Phi_{2,B}^\dagger \Phi_{2,B})^2 \notag \\
    &- \lambda_{3,B} (\Phi_{1,B}^\dagger \Phi_{1,B})(\Phi_{2,B}^\dagger \Phi_{2,B}) - \lambda_{4,B} (\Phi_{2,B}^\dagger \Phi_{1,B})(\Phi_{1,B}^\dagger \Phi_{2,B}) \notag \\
    &- \bigg\{ \Big( \frac{1}{2}\lambda_{5,B} \Phi_{1,B}^\dagger \Phi_{2,B} +  \lambda_{6,B} \Phi_{1,B}^\dagger \Phi_{1,B} + \lambda_{7,B} \Phi_{2,B}^\dagger \Phi_{2,B} \Big) \Phi_{1,B}^\dagger \Phi_{2,B} + \mathrm{h.c.} \bigg\},
    \label{eq:bare_lagrangian}
\end{align}
where the subscripts $B$ mean the bare quantities.
The two doublet fields are renormalized as~\cite{Hollik:1993cg}
\begin{align}
    \Phi_{a,B} = Z_{ab} \Phi_b,~(a,b=1,2)
\end{align}
where $Z_{ab} \in \mathbb{C}$. We define the renormalized couplings as~\cite{Hollik:1993cg}
\begin{alignat}{4}
    &Z_{\mu_1} \mu_1^2 \equiv \mu_{1,B}^2 |Z_{11}|^2, &~~&Z_{\mu_2} \mu_2^2 \equiv \mu_{2,B}^2 |Z_{22}|^2,&~~&Z_{\mu_3} \mu_3^2 \equiv \mu_{3,B}^2 Z_{11}^* Z_{22}, \notag \\
    &Z_{\lambda_1} \lambda_1 \equiv \lambda_{1,B} |Z_{11}|^4, &~~&Z_{\lambda_2} \lambda_2 \equiv \lambda_{2,B} |Z_{22}|^4, &~~&Z_{\lambda_3} \lambda_3 \equiv \lambda_{3,B} |Z_{11}|^2|Z_{22}|^2, \notag \\
    &Z_{\lambda_4} \lambda_4 \equiv \lambda_{4,B} |Z_{11}|^2|Z_{22}|^2, &~~&Z_{\lambda_5} \lambda_5 \equiv \lambda_{5,B} Z_{11}^{2*} Z_{22}^2,&~~& Z_{\lambda_6} \lambda_6 \equiv \lambda_{6,B}|Z_{11}|^2 Z_{11}^* Z_{22}, \notag \\
    &Z_{\lambda_7} \lambda_7 \equiv \lambda_{7,B}|Z_{22}|^2 Z_{11}^* Z_{22},
\end{alignat}
The one-loop counter terms are obtained by shifting the renormalization constants as $Z_{ab}\to \delta_{ab}+\delta Z_{ab}/2$, $Z_{\mu_{i}} \mu_{i}^2 \to \mu_{i}^2 + \delta \mu_{i}^2$ and $Z_{\lambda_{i}} \lambda_{i} \to \lambda_{i} + \delta \lambda_{i}$.

The renormalized fields are defined by 
\begin{align}
    \Phi_1 = 
    \begin{pmatrix}
        G^+ \\ \frac{1}{\sqrt{2}} (v + h_1 + iG^0)
    \end{pmatrix}, ~~~
    \Phi_2 = 
    \begin{pmatrix}
        H^+ \\ \frac{1}{\sqrt{2}} (h_2 + i h_3)
    \end{pmatrix},
\end{align}
where the renormalized VEV is written by 
\begin{align}
    v^{2} = \frac{2 \mu_1^2}{\lambda_1} = \frac{2 \mu_3^2}{\lambda_6}.
\end{align}

In this paper, we adopt a renormalization scheme used in the effective potential.
We parametrize the electrically neutral VEVs of the two doublet fields as 
\begin{align}
	\langle \Phi_1 \rangle = \frac{1}{\sqrt{2}}
	\begin{pmatrix}
	0 \\ \varphi_1
	\end{pmatrix}, ~~~~
	\langle \Phi_2 \rangle = \frac{1}{\sqrt{2}}
	\begin{pmatrix}
	0 \\   \varphi_2 + i\varphi_3
	\end{pmatrix},
\end{align}
where $\varphi_i \in \mathbb{R}$.
At the one-loop level, the effective potential at zero temperature is given by 
\begin{align}
	V^{\mathrm{eff}}_{T=0}(\varphi_1,\varphi_2,\varphi_3) = V_0 + V_{\mathrm{CW}} + V_{\mathrm{CT}},
    \label{eq:effective_potential}
\end{align}
where $V_0$, $V_{\mathrm{CW}}$ and $V_{\mathrm{CT}}$ are the tree-level potential, the Coleman--Weinberg (CW) potential~\cite{Coleman:1973jx} and the corrections from the counter term, respectively.
We calculate the CW potential in the Landau gauge.
The explicit form of them is given in refs.~\cite{Cline:2011mm,Enomoto:2022rrl}.

At the tree level, $\bm{\varphi}=(\varphi_1 ,\varphi_2, \varphi_3)=(v,0,0) \equiv \bm{\varphi}_{vac}$ is the minimum of the Higgs potential.
However, radiative corrections change the vacuum position and the curvature from those at the tree level.
We determine $V_{\mathrm{CT}}$ so that loop corrections do not change the vacuum position and the curvatures at the vacuum from the tree-level potential.
We impose nine renormalization conditions
\begin{align}
	&\left. \frac{\partial V^{\mathrm{eff}}_{T=0}}{\partial \varphi_i}\right|_{\bm{\varphi} = \bm{\varphi}_{vac}} = 0,~~~(i=1,2,3), \label{eq:EP_tadpole} \\
	&\left. \frac{\partial^2 V^{\mathrm{eff}}_{T=0}}{\partial \varphi_i \partial \varphi_j} \right|_{\bm{\varphi} = \bm{\varphi}_{vac}} = \mathcal{M}^2_{ij},~~~(i,j=1,2,3), \label{eq:EP_mass}
\end{align}
and the other parameters, which are not determined by these conditions, are renormalized by the $\overline{\mathrm{MS}}$ scheme.
We also determine $Z_{ab}$ in the $\mathrm{\overline{MS}}$ scheme.
We refer to this renormalization scheme as the \textit{Effective Potential (EP) scheme}.
We note that $Z_{ab}$ are irrelevant for the EDM calculations in the EP scheme.

From eqs.~\eqref{eq:EP_tadpole}-\eqref{eq:EP_mass}, we obtain the renormalization conditions for the one- and two-point functions as 
\begin{align}
    \widehat{\Gamma}_{i} &\equiv \Gamma_{i} +\delta \Gamma_{i} = 0, \label{eq:EP_onepoint} \\
    % \widehat{\Pi}_{ij}(p^2 = 0) &\equiv \Pi_{ij} (p^2 = 0) + \delta \Pi_{ij} = 0, \label{eq:EP_twopoint}
    \widehat{\Pi}_{ij}(0) &\equiv \Pi_{ij} (0) + \delta \Pi_{ij} = 0, \label{eq:EP_twopoint}
\end{align}
where $\widehat{\Gamma}_{i}$ and $\widehat{\Pi}_{ij} (p^2)$ are the renormalized one-point functions for $h_i$ and the two-point self energies for $h_i$ and $h_j$ with the external momentum $p^2$, respectively.
These renormalized functions are given by the sum of the 1PI diagrams $\Gamma_i$ ($\Pi_{ij}$) and its counter term $\delta \Gamma_i$ ($\delta \Pi_{ij}$).
Unless otherwise noted, the coupling constants shown in the following discussions are understood as the renormalized coupling in the EP scheme.

\subsection{Thermal corrections}

The effective potential in finite temperature is given by
\begin{align}
	V^{\mathrm{eff}}(\varphi_1,\varphi_2,\varphi_3;T) = V^{\mathrm{eff}}_{T=0} + V_T.
\end{align}
The thermal correction $V_T$ is given by~\cite{Dolan:1973qd}
\begin{align}
	V_T = \sum_{\substack{i = \mathrm{fermion} \\ ~~\mathrm{boson}}} (-1)^{s_i} \frac{n_i}{2\pi^2 \beta^4} \int_0^\infty dx~x^2 \log \left( 1 + (-1)^{s_i+1} \exp \left( -\sqrt{x^2+\beta^2 \tilde{m}_i^2} \right)  \right),
\end{align}
where $\beta = 1/T$, $n_i$ is the degree of freedom of a particle $i$, $s_i = 0$ for bosons and $s_i = 1$ for fermions, and $\tilde{m}^2_i$ is the field dependent mass.
We employ the Parwani resummation scheme \cite{Parwani:1991gq} to include leading-order thermal corrections.

\section{Scenario for electroweak baryogenesis \label{sec:EWBG}}

In this section, we specify our scenario for EWBG.
We discuss which rephasing invariants in eqs.~\eqref{eq:rephasing_invariant1} and \eqref{eq:rephasing_invariant2} are crucial for the BAU under the current experimental constraints.
We also discuss the EW phase transition in our scenario.

\subsection{CP-violating phase \label{sec:minimal_setup}}

Here, we investigate EWBG under the current experimental constraints.
We specify which rephasing invariants in eqs.~\eqref{eq:rephasing_invariant1} and \eqref{eq:rephasing_invariant2} are crucial for the BAU.

First, we discuss the constraints from the Higgs signal measurements.
By the current measurements at the LHC~\cite{ATLAS:2022vkf,CMS:2022dwd}, the $H_{1}$ couplings such as $H_1 ZZ$ and $H_1 WW$ are required to be consistent with the SM at $\mathcal{O}(10)~\%$ level.
This leads to the constraints on the mixing angles $\alpha_i$ among the neutral scalar bosons.
Since the $\lambda_6$ coupling causes non-zero $\alpha_i$, the invariants including $\lambda_6$ are unsuitable for the BAU.
Thus, we consider a scenario with $\lambda_6 = 0$ in the following discussions.
In this limit, there remain the following rephasing invariants expressing CP violations:
\begin{alignat}{4}
    &\mathrm{Im}[\lambda_5^* \lambda_7^2], ~~&
    &\mathrm{Im}[\lambda_5 (\rho^u)_{lm}^2], ~~& 
    &\mathrm{Im}[\lambda_5^* (\rho^{D})_{lm}^2], ~~&
    &\mathrm{Im}[\lambda_7 (\rho^u)_{lm}], \notag \\
    &\mathrm{Im}[\lambda_7^* (\rho^{D})_{lm}], ~~&
    &\mathrm{Im}[(\rho^u \rho^{D})_{lm}], ~~& 
    &\mathrm{Im}[(\rho^u \rho^{u \dagger})_{lm}], ~~& 
    &\mathrm{Im}[(\rho^D \rho^{D \dagger})_{lm}].
    \label{eq:rephasing_invariant3}
\end{alignat}

Second, we specify which Yukawa sector should involve CP violations for EWBG to work sufficiently.
In the following analysis, we calculate the BAU by using the semi-classical force method with the WKB approximation~\cite{Cline:2000nw,Fromme:2006wx,Joyce:1994fu,Joyce:1994zn,Joyce:1994zt,Cline:2020jre}.
In this method, the CPV source term in the transport equation for a fermion $f_l$ is given by~\cite{Cline:2020jre}
\begin{align}
    S_{f_l} = -\gamma v_w \big(m_{f_l}^2 \theta_{f_l}^\prime \big)^\prime Q_{f_l}^8 + \gamma v_w m_{f_l}^2 \theta_{f_l}^\prime \big(m_{f_l}^2 \big)^\prime Q_{f_l}^9,
    \label{eq:CPV_source}
\end{align}
where $\gamma = \sqrt{1 -v_w^2}$ with the velocity of the expanding bubble wall $v_w$, and $Q^{8,9}_{f_l}$ are functions of the mass of $f_l$.
We here have neglected the curvature of the expanding bubble, taking the plane wall approximation.
We have also assumed that $v_w$ is constant, and we will treat it as a free parameter.
The squared mass $m_{f_l}^2$ and its phase $\theta_{f_l}$ depend on the bubble profile.
The prime of them means the derivative with respect to the spacial radial coordinate of the bubble in the co-moving frame of the wall.

When we take $\lambda_6 = 0$ with $\chi = -\mathrm{arg}[\lambda_5]/2$, the local mass and the phase for the top quarks induced by $\rho_{tt}$ are given by~\cite{Cline:2011mm}
\begin{align}
	m_t^2 &= \frac{1}{2}\Big( y_t^2 \varphi_1^2 + |\rho_{tt}|^2(\varphi_2^2+\varphi_3^2) + 2y_t |\rho_{tt}| \varphi_1 \big(\varphi_2 \cos\theta_{tt} + \varphi_3\sin\theta_{tt} \big) \Big), \label{eq:local_mass_and_phase1} \\
	m_t^2 \theta_t^\prime &= \frac{1}{2}~ \Bigg\{ y_t |\rho_{tt}| \Big(  (\varphi_3 \varphi_1^\prime - \varphi_1 \varphi_3^\prime) \cos\theta_{tt} + (\varphi_1 \varphi_2^\prime - \varphi_2 \varphi_1^\prime ) \sin\theta_{tt} \Big) \notag \\
    &+ |\rho_{tt}|^2 ( \varphi_3 \varphi_2^\prime  - \varphi_2 \varphi_3^\prime ) \Bigg\} +\frac{m_t^2}{\varphi_1^2 + \varphi_2^2+\varphi_3^2}(\varphi_3 \varphi_2^\prime  - \varphi_2 \varphi_3^\prime ),
    \label{eq:local_mass_and_phase2}
\end{align}
where $\theta_{tt} = \mathrm{arg}[\rho_{tt}]$.
On the right-hand side of eq.~\eqref{eq:local_mass_and_phase2}, the terms depending on $y_t|\rho_{tt}|$ dominate the contributions to $m_t^2 \theta_t^\prime$.
Hence, the CPV source term is approximately proportional to $y_t|\rho_{tt}|$.
This can be understood as follows.
In the single step phase transition, the vacuum changes from $\bm{\varphi} \simeq (0,0,0)$ to $\bm{\varphi} \simeq (v_n, 0, 0)$, and the deviations of $\varphi_2$ and $\varphi_3$ from zero are suppressed by the positive curvature of the potential in the directions of $\varphi_{2}$ and $\varphi_3$.
As a result, the bounce solution for $\varphi_1$ is typically much larger than that for $\varphi_2$ and $\varphi_3$, and the first term gives the leading contribution.

Compared to the top quarks, the CPV source terms in eq.~\eqref{eq:CPV_source} for the other light fermions are suppressed by $y_{f_l} |(\rho^f)_{ll}| / y_t |\rho_{tt}|$, and they are, in general, sub-leading for generating the BAU.
In addition, it is known that the effects of the off-diagonal FCNC couplings on the CPV source terms are sub-leading~\cite{Kanemura:2023juv}.
After taking $(\rho^f)_{lm}=0$ except for $\rho_{tt}$, the candidates are reduced to be
\begin{align}
    \mathrm{Im}[\lambda_5^* \lambda_7^2],\quad \mathrm{Im}[\lambda_5 \rho_{tt}^2],\quad \mathrm{Im}[\lambda_7 \rho_{tt}].
    \label{eq:rephasing_invariant4}
\end{align}

Third, if we take $\lambda_6 = \lambda_7 = 0$, the tree-level potential $V_{0}$ is symmetric under the $\mathbb{Z}_2$ transformation $\Phi_2 \to -\Phi_2$.
This potential is the same as the one of the Inert doublet model~\cite{Deshpande:1977rw}.
Due to the Yukawa interaction, this $\mathbb{Z}_2$ symmetry is violated by the one-loop effects in the effective potential $V^{\mathrm{eff}}(\varphi_1,\varphi_2,\varphi_3;T)$.
However, the effective potential is still almost symmetric under the transformation $\varphi_{2,3} \to - \varphi_{2,3}$ because the one-loop corrections are too small to change the global structure of the potential.
As a result, the bounce solutions of $\varphi_2$ and $\varphi_3$ are almost zero, and the baryon asymmetry cannot be large enough.
In other words, among the three invariants in eq.~\eqref{eq:rephasing_invariant4}, the terms with non-zero $\lambda_7$ are necessary to generate the sufficient BAU.
In addition, since the CPV source is approximately proportional to $|\rho_{tt}|$, non-zero $\rho_{tt}$ is also needed.
Consequently, the scenario with $\rho_{tt} = 0$ or $\lambda_7 = 0$ is not successful, and $\mathrm{Im}[\lambda_7 \rho_{tt}]$ is essential for the BAU.
In the following, we take $\lambda_5 = 0$ as we would like to minimize the parameter space to focus on the CP violation relevant to the BAU and examine its impacts on the EDMs.

Fourth, as for the parameters irrelevant to the rephasing invariants, we need to satisfy the constraint from the $T$ parameter~\cite{Peskin:1990zt,Peskin:1991sw,Lytel:1980zh,Toussaint:1978zm,Pomarol:1993mu,Haber:2010bw}.
The additional scalar bosons contribute to the self-energy of the W and Z bosons.
The correction to the $T$ parameter in the 2HDM, $\Delta T$, is proportional to the squared mass difference of $H^\pm$ and $H_2$ or $H_3$~\cite{Pomarol:1993mu,Haber:2010bw}.
With $\lambda_4 = \lambda_5$ and $\lambda_6 = 0$ on the $\chi = -\mathrm{arg}[\lambda_5]/2$ basis, we obtain $\Delta T \simeq 0$ at the one-loop level, so that we take $\lambda_4 = \lambda_5$ to avoid the $T$ parameter constraint.

Finally, let us specify our scenario for EWBG. As we have discussed, $\rho_{tt}$ and $\lambda_{7}$ are important to realize EWBG successfully. 
Assuming the irrelevant parameters vanished for the minimality, the model parameters of interest are shown as
\begin{align}
    \mathrm{Im}[\lambda_7 \rho_{tt}] \neq 0 \qc
    \lambda_4 = \lambda_5 = \lambda_6 = 0,~~\mathrm{and}~~\rho^f = \bm{0},~~\mathrm{except~for}~\rho_{tt}.
    \label{eq:minimal_setup}
\end{align}
We note that $\lambda_6 = 0$ with $\chi = -\mathrm{arg}[\lambda_5]/2$ leads to $\mathcal{R}_{ij}=\delta_{ij}$.
In addition, due to $\lambda_{4}=\lambda_{5}=0$, the additional scalar bosons $H^\pm, H_2$ and $H_3$ are degenerate in their masses.
We define $m_{\Phi}^2 \equiv m_{H^\pm}^2 = m_{H_2}^2 = m_{H_3}^2$, where $m_{\Phi}^2 = M^2 + \frac{1}{2} \lambda_3 v^2$.
We treat $m_{\Phi},\ M,\ \lambda_2,\ \lambda_7$, and $\rho_{tt}$ as independent parameters.
As we will discuss in section~\ref{sec:BAU_and_EDM}, in our scenario, the generated BAU and the EDMs are strongly correlated because there is only one rephasing invariant, $\mathrm{Im}[\lambda_7 \rho_{tt}]$.

\subsection{Electroweak phase transition \label{sec:EWPT}}

We here discuss the EW phase transition in our scenario.
The EW phase transition and the bubble nucleation are calculated based on $V^{\mathrm{eff}}(\varphi_1,\varphi_2,\varphi_3;T)$.
The nucleation temperature $T=T_n$ is defined by $\Gamma (T_n) / H(T_n)^4 = 1$, where $H(T)$ is the Hubble parameter and $\Gamma(T)$ is the probability of the vacuum transition per unit time and volume.
We define $v_n$ as the VEV at $T=T_n$.
We numerically solve the bounce equations and obtain the bubble profiles by using \texttt{CosmoTransitions}~\cite{Wainwright:2011kj}.

\begin{figure}[t]
    \centering
    \includegraphics[width=1\linewidth]{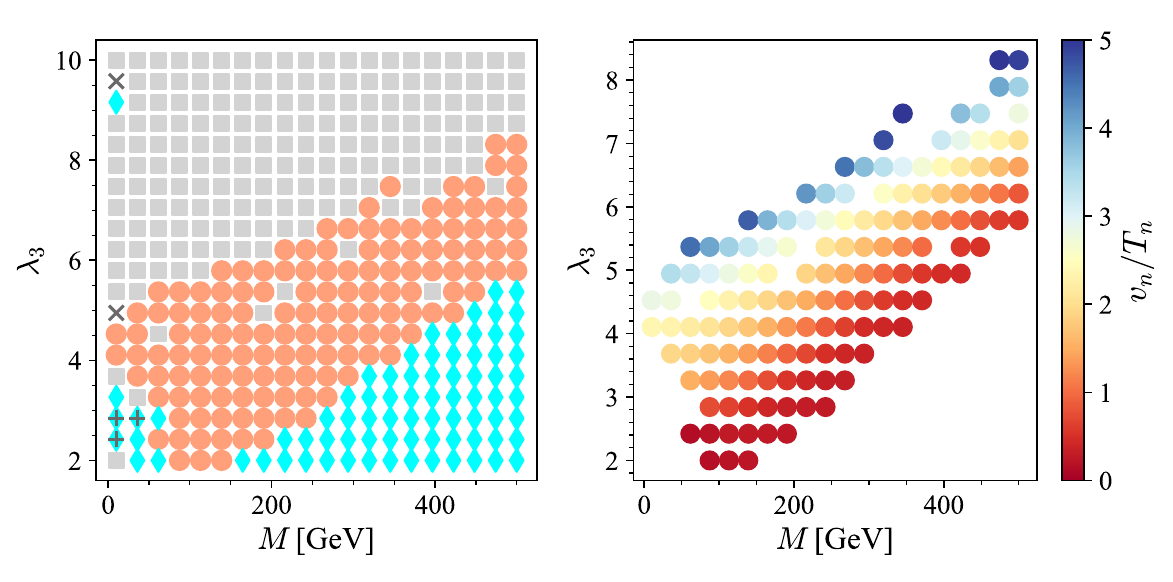}
    \caption{Left: EW phase transition in our scenario. 
    See the main text for the explanation of each color point.  
    EWBG could be realized in the orange-filled encircled points.
    Right: The value of $v_n / T_n$ in the orange-filled encircled points in the left panel.}
    \label{fig:EWPT}
\end{figure}
In the left panel of figure~\ref{fig:EWPT}, the behavior of the EW phase transition is shown in the $\lambda_3$--$M$ plane.
We have set $\lambda_2 = 0.2$, $\lambda_7 = 1e^{-i\pi/4}$ and $\rho_{tt}= 0.1e^{-i\pi/4}$.
We classify the fates of the vacuum as follows:
\begin{description}
    \item[(a) Orange filled circle:]
    The phase transition from phase $A$ to phase $B$ ($A\to B$) is first order, $B$ is the vacuum at $T=0$, $B$ is the EW vacuum, 
    and the VEV in $A$ is \textit{lower} than 1~GeV.
    \item[(b) Gray $\bm{+}$:] The phase transition $A \to B$ is first order, $B$ is the vacuum at $T=0$, \\ $B$ is the EW vacuum, and the VEV in $A$ is \textit{larger} than 1~GeV.
    \item[(c) Gray $\bm{\times}$:] The phase transition $A \to B$ is first order, $B$ is the vacuum at $T=0$, \\ and $B$ is \textit{not} the EW vacuum.
    \item[(d) Black dot:] The phase transition $A \to B$ is first order, and $B$ is \textit{not} the vacuum at $T=0$.
    \item[(e) Blue diamond:] The phase transition $A \to B$ is second order.
    \item[(f) Gray box:] Other cases including $\Gamma / H^4 < 1$.
\end{description}
EWBG works successfully in case (a), in which we calculate the BAU.
We note that, in figure~\ref{fig:EWPT}, there are no phase transitions classified in case (d).
The overlapped points mean the case when the multi-step phase transition happens.
For the strongly first-order EW phase transition in the 2HDM, the non-decoupling quantum effects of the additional scalar bosons play important roles~\cite{Kanemura:2004ch}.
With our setup given in eq.~\eqref{eq:minimal_setup}, the non-decoupling effects are caused by $\lambda_3$.
As one can see in the figure~\ref{fig:EWPT}, for the first-order EW phase transition, a condition $m_{\Phi}^2 \simeq \frac{1}{2} \lambda_3 v^2 \gg M^2$ with a relatively large $\lambda_3$ is needed.
As $\lambda_3$ grows, the phase transition becomes first order by the non-decoupling quantum effects of the additional scalar bosons.
With even larger $\lambda_3$, the decay rate $\Gamma$ does not exceed the fourth power of the Hubble parameter (the case shown by the gray boxes), and the EW phase transition has not been completed until now.

In the right panel of figure~\ref{fig:EWPT}, we show the behavior of $v_n / T_n$ in case (a).
The color gradient indicates the value of $v_n / T_n$.
For EWBG to work successfully, the sphaleron process must be immediately decoupled in the broken phase, which requires the condition $v_n / T_n \gtrsim 1$~\cite{Moore:1998swa}.
We adopt $v_n / T_n > 1$ as the criterion of the strongly first-order EW phase transition.

\section{Top-quark (chromo-)EDM \label{sec:EDM}}
In this section, we discuss the (C)EDM of the top quark in our scenario.
They induce the neutron, proton, and electron EDMs via radiative corrections.
We find that there are no one-loop contributions to the top-quark (C)EDM.
We provide formulae for the leading two-loop corrections to the (C)EDM induced by $\Im[\lambda_{7}\rho_{tt}]$.

We define the relevant CPV effective operators as
\footnote{
In general, we also have the $\theta$ term 
\begin{align*}
    \mathcal{L}_{\mathrm{CPV}} = \theta \frac{g_S^2}{32 \pi^2} \tilde{G}^a_{\mu \nu} G^{a \mu \nu}.
\end{align*}
This operator induces the (C)EDMs, and $|\theta|$ must be smaller than $\mathcal{O}(10^{-10})$ from the current bound~\cite{Abel:2020pzs}.
It is expected that the top-quark CEDM discussed below induces large $|\theta|$.
The problem of the smallness of $|\theta|$ is known as the strong CP problem.
In this paper, we do not further enter this problem by assuming some mechanisms working to suppress this term, e.g., the Peccei-Quinn mechanism~\cite{Peccei:1977hh,Peccei:1977ur}.
}
\begin{align}
    \mathcal{L}_{\mathrm{CPV}} = -\frac{1}{2} d_\psi \overline{\psi} \sigma_{\mu \nu} i \gamma^5 \psi F^{\mu \nu} -\frac{1}{2} g_S \tilde{d}_q \overline{q} \sigma_{\mu \nu} i \gamma^5 T^a q G^{a \mu \nu} + \frac{1}{3}w f_{abc} G_{\mu \nu}^a \tilde{G}^{b \nu \rho} G^{c \mu}_\rho,
\end{align}
where $\psi$ is the Dirac spinor of the lepton $l$ or the quark $q$, and $\sigma_{\mu \nu} = \frac{i}{2}[\gamma^\mu, \gamma^\nu]$.
The strong coupling constant and the generator of $\mathrm{SU}(3)_C$ are denoted by $g_S$ and $T^a = \lambda^a /2$ ($a = 1,...,8$), respectively, where $\lambda^a$ are the Gell-Mann matrices.
The field strength tensors for the photon and gluon are $F^{\mu \nu}$ and $G^{a \mu \nu}$, respectively, and its self-dual is defined by $\tilde{G}^a_{\mu \nu} \equiv \frac{1}{2} \epsilon_{\mu \nu \rho \sigma} G^{a \rho \sigma}$ with $\epsilon_{0123} = +1$.
The first, second, and third terms correspond to the EDM, CEDM, and Weinberg operators~\cite{Weinberg:1989dx, Dicus:1989va}, respectively. 

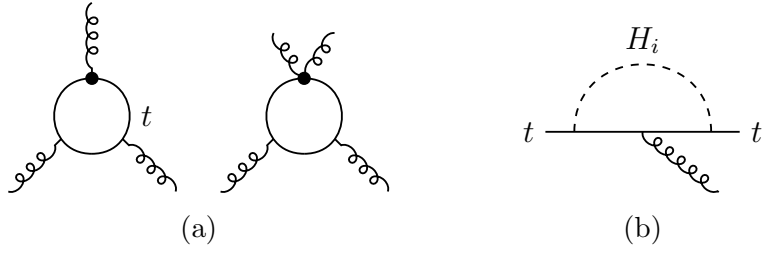
\begin{figure}[t]
    \centering
    \setlength{\feynhandlinesize}{0.7pt}
    \setlength{\feynhandarrowsize}{5pt}
    \begin{minipage}[t]{0.4\columnwidth}
    \centering
    \begin{tikzpicture}
        \begin{feynhand}
        \vertex (a) at (0,0); \vertex (b) at (-0,-1);
        \vertex (c) at (0.4,-0.8); \vertex (d) at (0,1);
        \vertex (e) at (1.1,-1.5); \vertex (f) at (-0.4,-0.8);
        \vertex (g) at (-1.1,-1.5);
        \vertex [dot] (h) at (0,0) {};
        \propag[glu] (d) to (a); \propag[glu] (e) to (c);
        \propag[glu] (g) to (f);
        \propag[plain] (a) to [half left, looseness=1.7, edge label=$t$] (b);
        \propag[plain] (b) to [half left, looseness=1.7] (a);
        \end{feynhand}
    \end{tikzpicture} ~~ 
    \begin{tikzpicture}
        \begin{feynhand}
        \vertex (a) at (0,0); \vertex (b) at (-0,-1);
        \vertex (c) at (0.4,-0.8); \vertex (d) at (0.4,0.6);
        \vertex (e) at (1.1,-1.5); \vertex (f) at (-0.4,-0.8);
        \vertex (g) at (-1.1,-1.5);
        \vertex [dot] (h) at (0,0) {};
        \vertex (i) at (-0.4,0.6);
        \propag[glu] (d) to (a); \propag[glu] (e) to (c);
        \propag[glu] (g) to (f);
        \propag[plain] (a) to [half left, looseness=1.7] (b);
        \propag[plain] (b) to [half left, looseness=1.7] (a);
        \propag[glu] (i) to (a);
        \end{feynhand}
    \end{tikzpicture}
    \subcaption{}
  \end{minipage}
  ~~
  \begin{minipage}[t]{0.3\columnwidth}
    \centering
    \begin{tikzpicture}
        \begin{feynhand}
        \vertex (a) at (0,0){$t$}; \vertex[particle] (b) at (3,0){$t$};
        \vertex (c) at (0.6,0); \vertex (d) at (2.4,0);
        \vertex (e) at (1.5,0);\vertex (f) at (2.5,-0.8);
        \propag[plain] (a) to (b);
        \propag[sca] (c) to [half left, looseness=1.7, edge label=$H_i$] (d);
        \propag[glu] (e) to (f);
        \end{feynhand}
    \end{tikzpicture}
    \subcaption{}
  \end{minipage}
  \caption{(a) Threshold corrections to the Weinberg operator at the top-quark mass.
  (b) One-loop CEDM diagram for the top quark.}
  \label{fig:CEDMdiagram1}
\end{figure}

\subsection{Neutron and proton EDMs induced by top-quark CEDM \label{sec:npEDM}}

The neutron and proton EDM are given by the QCD sum rules~\cite{Demir:2002gg, Pospelov:2005pr, Fuyuto:2013gla, Haisch:2019bml, Kaneta:2023wrl}
\begin{align}
    &d_n = 0.73 d_d - 0.18 d_u + e(0.20 \tilde{d}_d + 0.10 \tilde{d}_u) + 23 \times 10^{-3} ~\mathrm{GeV} ~e w, \notag \\
    &d_p = 0.73 d_u - 0.18 d_d - e(0.40 \tilde{d}_u + 0.049 \tilde{d}_d)- 33 \times 10^{-3} ~\mathrm{GeV} ~e w,
\end{align}
where $e$ is the electromagnetic coupling constant.

In our scenario, the (C)EDMs of the light quarks and the Weinberg operator are caused by the top-quark CEDM.
Below the mass threshold of the top quarks, the Weinberg operator is induced~\cite{Braaten:1990gq,Chang:1991ry,Kamenik:2011dk}, as shown in the left panel of figure.~\ref{fig:CEDMdiagram1}.
The threshold correction is shown as~\cite{Kamenik:2011dk}
\begin{align}
    \delta w^{(t)} / g_S = \frac{g_S^2}{32 \pi^2} \frac{\tilde{d}_t}{m_t},
\end{align}
at the top mass scale $m_t$.
It contributes to the other CPV operators via radiative corrections.
The renormalization group flow is found in refs.~\cite{Degrassi:2005zd,Hisano:2012cc,Jung:2013hka,Brod:2018pli}. 
From the top-quark CEDM, they are obtained as
\begin{align}
    &d_u = 1.8 \times 10^{-9} ~ e ~ \tilde{d}_t, ~~~ d_d = -2.0 \times 10^{-9} ~ e ~ \tilde{d}_t, \notag \\
    &\tilde{d}_u = -8.0 \times 10^{-9} ~ \tilde{d}_t, ~~~ \tilde{d}_d = -1.7 \times 10^{-8} ~ \tilde{d}_t, \notag \\
    &w = - 1.4 \times 10^{-5} ~\mathrm{GeV}^{-1} ~ \tilde{d}_t,
    \label{eq:dtcinduce}
\end{align}
at the hadronization scale $\mu_H = 2$~GeV.

\subsubsection{One-loop CEDM}

Let us discuss the top-quark CEDM at the one-loop level. 
From the diagram on the right panel of figure~\ref{fig:CEDMdiagram1}, we obtain
\begin{align}
    \tilde{d}_t^{(1)} = - \frac{m_t}{16 \pi^2} \sum_{i=1}^3 \Big( y_t \mathcal{R}_{1i} + \rho_{tt}^R \mathcal{R}_{2i} +\rho_{tt}^I \mathcal{R}_{3i} \Big) \Big( \rho_{tt}^I \mathcal{R}_{2i} - \rho_{tt}^R \mathcal{R}_{3i} \Big) C_{11}[H_i,t,t],
    \label{eq:topCEDM1}
\end{align}
where the loop function $C_{11}[H_i,t,t]$ is given in appendix~\ref{sec:Loopfunc}.
We note that this expression is general, and the conditions in eq.~\eqref{eq:minimal_setup} have not been imposed.
\begin{figure}
    \centering
    \includegraphics[width=0.7\linewidth]{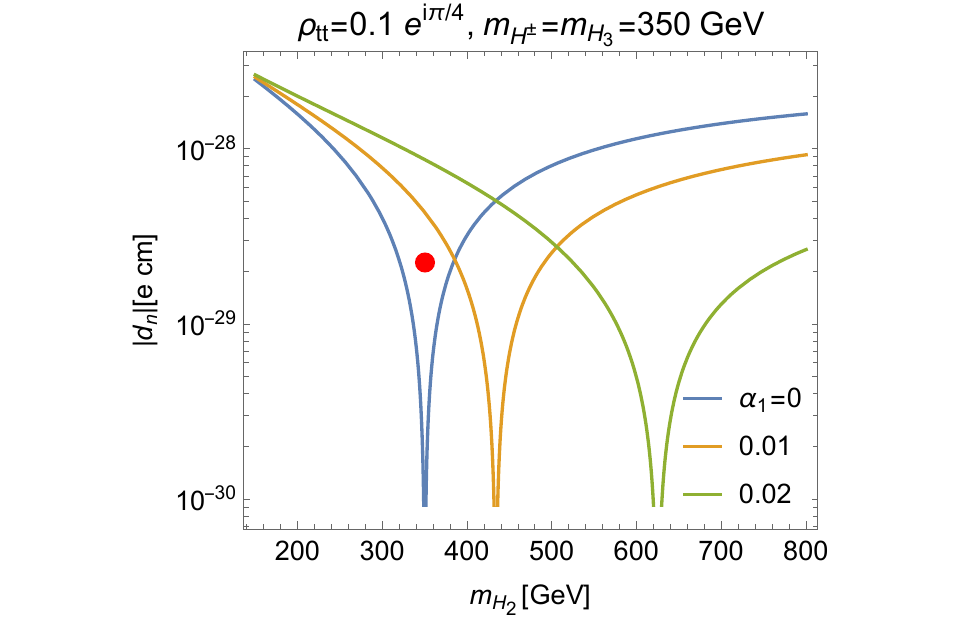}
    \caption{The neutron EDM induced by the top-quark CEDM.
    The solid blue, orange, and green curves are caused by the one-loop contributions in the cases with $\alpha_1 = 0,0.01$ and $0.02$, respectively.
    The red point at $m_{H_2} = m_{H^\pm} = m_{H_3}= 350~\mathrm{GeV}$ and $\alpha_1 =0$ shows $|d_n|=2.2 \times 10^{-29}~e~\mathrm{cm}$ induced by the two-loop contributions.
    At the point, we have taken $\lambda_7 = 1 e^{i \pi /4}$ and $M=30~\mathrm{GeV}$.}
    \label{fig:nEDM_1loop}
\end{figure}
In figure~\ref{fig:nEDM_1loop}, we show $|d_n|$ as a function of $m_{H_2}$.
We take $m_{H^\pm} = m_{H_3} = 350$~GeV, $\rho_{tt} =0.1e^{i \pi/4}$ and $\alpha_2 = 0$ with $\chi = -\mathrm{arg}[\lambda_5]/2$.
The blue, orange, and green curves correspond to the cases of $\alpha_1 = 0$, $0.01$, and $0.02$, respectively.
From the blue curve with $\alpha_1 = 0$, we can see that the neutron EDM vanishes at the point with $m_{H_2} = m_{H_3}$.
This is because $\tilde{d}_t^{(1)}$ is caused by $\mathrm{Im}[\lambda_5 \rho_{tt}^2]$ or $\mathrm{Im}[\lambda_6 \rho_{tt}]$ and both vanish at this point.
For the cases of $\alpha_1 \neq 0$, the neutron EDM vanishes at the point with $m_{H_2} \neq m_{H_3}$.
This is because $|d_n| = 0$ is realized by a cancellation between the EDM contributions from $\mathrm{Im}[\lambda_5 \rho_{tt}^2]$ and $\mathrm{Im}[\lambda_6 \rho_{tt}]$.
Since $\alpha_1 \neq 0$ leads to non-zero $\lambda_{6}$, the cancellation occurs at the point with non-zero $\lambda_5$, which leads to $m_{H_2} \neq m_{H_3}$.

Our scenario predicts $\tilde{d}_t^{(1)} =0$ because $\lambda_{5}=\lambda_{6}=0$ is assumed (see eq.~\eqref{eq:minimal_setup}).
Thus, both neutron and proton EDMs are also zero. 
In the following, we show that the leading contributions appear at the two-loop level.
The red point in figure~\ref{fig:nEDM_1loop} corresponds to them, which are explained in the following.

\subsubsection{Two-loop CEDM}

\begin{figure}[t]
    \centering
    \setlength{\feynhandlinesize}{0.7pt}
    \setlength{\feynhandarrowsize}{4pt}
    \begin{minipage}[t]{0.4\columnwidth}
    \centering
    \scalebox{1.2}{
    \begin{tikzpicture}
        \begin{feynhand}
        \vertex (a) at (-1,0); \vertex (b) at (1,0);
        \vertex (c) at (-0.4,0); \vertex (d) at (0.4,0);
        \vertex (e) at (-0.35,0.85); \vertex (f) at (0.35,0.85);
        \vertex (g) at (0,0.5);
        \vertex (h) at (0,1.2);
        \vertex (i) at (-0.6,0.27) {\scriptsize{$\Phi$}};
        \vertex (j) at (1,0.27) {\scriptsize{$\Phi_{\mathrm{SM}},V$}};
        \vertex (k) at (0,0.7) {\scriptsize{$\Phi$}};
        \vertex (l) at (0,1.4) {\scriptsize{$\Phi$}};
        \vertex (m) at (0,0); \vertex (n) at (0.7,-0.5);
        \propag[plain] (a) to (b); \propag[plain] (c) to (e);
        \propag[plain] (d) to (f);
        \propag[plain] (g) to [half left, looseness=1.7] (h);
        \propag[plain] (h) to [half left, looseness=1.7] (g);
        \propag[glu] (n) to (m);
        \end{feynhand}
    \end{tikzpicture}
    \begin{tikzpicture}
        \begin{feynhand}
        \vertex (a) at (-1,0); \vertex (b) at (1,0);
        \vertex (c) at (-0.6,0); \vertex (d) at (0.6,0);
        \vertex (e) at (0,0.5); \vertex (f) at (0,0.5);
        \vertex (g) at (0,0.5);
        \vertex (h) at (0,1.2);
        \vertex (i) at (-0.6,0.27) {\scriptsize{$\Phi$}};
        \vertex (j) at (0.8,0.27) {\scriptsize{$\Phi_{\mathrm{SM}}$}};
        \vertex (l) at (0,1.4) {\scriptsize{$\Phi$}};
        \vertex (m) at (0,0); \vertex (n) at (0.7,-0.5);
        \propag[plain] (a) to (b); \propag[plain] (c) to (e);
        \propag[plain] (d) to (f);
        \propag[plain] (g) to [half left, looseness=1.7] (h);
        \propag[plain] (h) to [half left, looseness=1.7] (g);
        \propag[glu] (n) to (m);
        \end{feynhand}
    \end{tikzpicture}
    }
    \subcaption*{(A) $\mathcal{O}(\lambda_7 \rho_{tt})$ contributions.}
  \end{minipage}
  ~~
  \begin{minipage}[t]{0.5\columnwidth}
    \centering
    \scalebox{1.2}{
    \begin{tikzpicture}
        \begin{feynhand}
        \vertex (a) at (-1,0); \vertex (b) at (1,0);
        \vertex (c) at (-0.4,0); \vertex (d) at (0.4,0);
        \vertex (e) at (-0.35,0.85); \vertex (f) at (0.35,0.85);
        \vertex (g) at (0,0.5);
        \vertex (h) at (0,1.2);
        \vertex (i) at (-0.6,0.27) {\scriptsize{$\Phi$}};
        \vertex (j) at (0.6,0.27) {\scriptsize{$\Phi$}};
        \vertex (k) at (0,0.7) {\scriptsize{$\Phi$}};
        \vertex (l) at (0,1.4) {\scriptsize{$\Phi$}};
        \vertex (m) at (0,0); \vertex (n) at (0.7,-0.5);
        \propag[plain] (a) to (b); \propag[plain] (c) to (e);
        \propag[plain] (d) to (f);
        \propag[plain] (g) to [half left, looseness=1.7] (h);
        \propag[plain] (h) to [half left, looseness=1.7] (g);
        \propag[glu] (n) to (m);
        \end{feynhand}
    \end{tikzpicture} 
    \begin{tikzpicture}
        \begin{feynhand}
        \vertex (a) at (-1,0); \vertex (b) at (1,0);
        \vertex (c) at (-0.6,0); \vertex (d) at (0.6,0);
        \vertex (e) at (0,0.5); \vertex (f) at (0,0.5);
        \vertex (g) at (0,0.5);
        \vertex (h) at (0,1.2);
        \vertex (i) at (-0.6,0.27) {\scriptsize{$\Phi$}};
        \vertex (j) at (0.6,0.27) {\scriptsize{$\Phi$}};
        \vertex (l) at (0,1.4) {\scriptsize{$\Phi$}};
        \vertex (m) at (0,0); \vertex (n) at (0.7,-0.5);
        \propag[plain] (a) to (b); \propag[plain] (c) to (e);
        \propag[plain] (d) to (f);
        \propag[plain] (g) to [half left, looseness=1.7] (h);
        \propag[plain] (h) to [half left, looseness=1.7] (g);
        \propag[glu] (n) to (m);
        \end{feynhand}
    \end{tikzpicture} 
    \begin{tikzpicture}
        \begin{feynhand}
        \vertex (a) at (-1,0); \vertex (b) at (1,0);
        \vertex (c) at (-0.6,0); \vertex (d) at (0.6,0);
        \vertex (e) at (0,0); \vertex (f) at (0,1);
        \vertex (g) at (-0.15,0.27) {\scriptsize{$\Phi$}};
        \vertex (i) at (-0.6,0.27) {\scriptsize{$\Phi$}};
        \vertex (j) at (0.6,0.27) {\scriptsize{$\Phi$}};
        \vertex (m) at (0,-0.2); \vertex (n) at (0.7,-0.6);
        \propag[plain] (a) to (b); \propag[plain] (c) to (f);
        \propag[plain] (d) to (f);
        \propag[plain] (e) to (f);
        \propag[glu] (n) to (m);
        \end{feynhand}
    \end{tikzpicture}
    }
    \subcaption*{(B) $\mathcal{O}(\lambda_7^2 \rho_{tt}^2)$ contributions.}
  \end{minipage}
  \caption{Topologies of two-loop diagrams relevant for the top-quark CEDM.}
  \label{fig:CEDMdiagram2}
\end{figure}
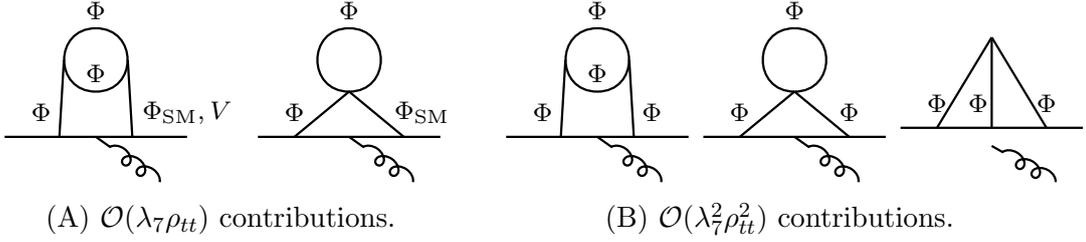
\begin{figure}[t]
    \centering
    \setlength{\feynhandlinesize}{0.7pt}
    \setlength{\feynhandarrowsize}{4pt}
    \begin{minipage}[t]{0.25\columnwidth}
    \centering
    \begin{tikzpicture}
        \begin{feynhand}
        \vertex (a) at (0,0){$t$}; \vertex[particle] (b) at (3,0){$t$};
        \vertex (c) at (0.6,0); \vertex (d) at (2.4,0);
        \vertex [ringblob] (g) at (1.5,1.0) {\footnotesize{$\widehat{\Pi}_{12,13}^{\lambda_7}$}};
        \vertex (e) at (1.5,0);\vertex (f) at (2.5,-0.7);
        \propag[plain] (a) to (b);
        \propag[sca] (c) to [quarter left, looseness=0.7, edge label=$H_1$] (g);
        \propag[sca] (g) to [quarter left, looseness=0.7, edge label=$H_{2,3}$] (d);
        \propag[glu] (f) to (e);
        \end{feynhand}
    \end{tikzpicture}
  \end{minipage}
  ~~
  \begin{minipage}[t]{0.6\columnwidth}
    \centering
    \begin{tikzpicture}
        \begin{feynhand}
        \vertex (e) at (-1.3,0){$\Pi_{12,13}^{\lambda_7}= $};
        \vertex (a) at (0,0){$H_1$}; \vertex (b) at (0.8,0);
        \vertex (c) at (2,0); \vertex (d) at (2.8,0){$H_{2,3}$};
        \propag[sca] (a) to (b);
        \propag[sca] (b) to [half left, looseness=1.5, edge label=$\Phi$] (c);
        \propag[sca] (c) to [half left, looseness=1.5, edge label=$\Phi$] (b);
        \propag[sca] (c) to (d);
        \end{feynhand}
    \end{tikzpicture}
    \begin{tikzpicture}[baseline=-6.5mm]
        \begin{feynhand}
        \vertex (a) at (-1,0){$H_1$}; \vertex (b) at (1.2,0){$H_{2,3}$};
        \vertex (d) at (0 ,0);
        \vertex (e) at (0 ,1);
        \vertex (f) at (0,1.3) {$\Phi$};
        \propag[sca] (a) to (b);
        \propag[sca] (d) to [half left, looseness=1.6] (e);
        \propag[sca] (e) to [half left, looseness=1.6] (d);
        \end{feynhand}
    \end{tikzpicture}
  \end{minipage}
  \caption{Two-loop diagrams for the top-quark CEDM.}
  \label{fig:CEDMdiagram3}
\end{figure}
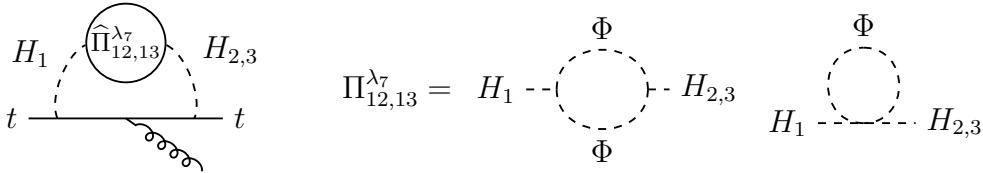

In our scenario, there is only one rephasing invariant, $\mathrm{Im}[\lambda_7 \rho_{tt}]$. This causes the top-quark CEDM at the two-loop level. 
In figure~\ref{fig:CEDMdiagram2}, we show the diagrams classified by the order of $\lambda_{7}\rho_{tt}$.
We have denoted the additional scalar bosons as $\Phi = H^\pm,H_2,H_3$, the SM scalar bosons as $\Phi_{\mathrm{SM}} = H_1, G^0, G^\pm$ and the gauge bosons as $V = W,Z,\gamma$.
Among the two-loop diagrams, type (A) diagrams are relevant because $\abs{\lambda_{7}\rho_{tt}}<1$ is required by the theoretical and experimental constraints.
In addition, among them, the additional scalar contributions are relevant compared with the gauge boson ones because large $\lambda_3$ is required by EWBG.
Thus, we focus on the diagrams in figure~\ref{fig:CEDMdiagram3}, which give the leading two-loop contributions.
We note that there are diagrams with $\Phi=H^{\pm}$ and $\Phi_{\mathrm{SM}}=G^{\pm}$ in (A).
However, they are subleading compared with the neutral scalar contributions due to the suppression by $m_b / m_t$.
The diagrams in figure~\ref{fig:CEDMdiagram3} include UV divergences, and we need to renormalize them. In the EP scheme discussed in section~\ref{sec:renormalization}, the counter terms for the one-point functions are obtained as
\begin{align}
    &\delta \Gamma_1 = \delta \mu_1^2 - \frac{1}{2} \delta \lambda_1 v^2 = -\frac{\Gamma_{H_1}}{v}, \notag \\
    &\delta \Gamma_2 = \mathrm{Re}\Big[ \delta \mu_3^2 - \frac{1}{2} \delta \lambda_6 v^2 \Big] = -\frac{\Gamma_{H_2}}{v}, \notag \\
    &\delta \Gamma_3 = \mathrm{Im}\Big[ \delta \mu_3^2 - \frac{1}{2} \delta \lambda_6 v^2 \Big] = \frac{\Gamma_{H_3}}{v}.
    \label{eq:tadpole_CT}
\end{align}
We note that the contributions from diagrams with the tadpole or the NG bosons become zero due to the tadpole conditions in eq.~(\ref{eq:EP_onepoint}).

What we are focusing on is the CP-odd EDM operator so that only diagrams involving $\lambda_7$ are relevant.
To calculate the top-quark CEDM, we need the renormalized mixed self energies involving $\lambda_7$ 
In the following, we denote them as $\widehat{\Pi}_{12}^{\lambda_7}$ and $\widehat{\Pi}_{13}^{\lambda_7}$.
From eq.~\eqref{eq:EP_twopoint}, the counter terms $\delta \Pi_{12}$ and $\delta \Pi_{13}$ are given by
\begin{align}
    \delta \Pi_{1 2} & = \mathrm{Re} \Big[ \delta \mu_3^2 - \frac{3}{2} \delta \lambda_6 v^2 \Big] = -\frac{\Gamma_{H_2}}{v} - \delta \lambda_6^R v^2, \notag \\
    \delta \Pi_{1 3} &= -\mathrm{Im} \Big[ \delta \mu_3^2 - \frac{3}{2} \delta \lambda_6 v^2 \Big] = -\frac{\Gamma_{H_3}}{v} + \delta \lambda_6^I v^2.
    \label{eq:Piij_CT}
\end{align}
In the second equality, we have used the tadpole conditions given in eq.~\eqref{eq:tadpole_CT}.
We obtain the relevant 1PI contributions as
\begin{align}
    \Pi_{1 2}^{\lambda_7} (p^2) &= \frac{3}{16 \pi^2} \lambda_3 \lambda_7^R v^2 B_0 [p^2; m_{\Phi}^2, m_{\Phi}^2] + \frac{3}{16\pi^2} \lambda_7^R A_0[m_{\Phi}^2], \notag \\
    \Pi_{1 3}^{\lambda_7} (p^2) &= -\frac{3}{16 \pi^2} \lambda_3 \lambda_7^I v^2 B_0 [p^2; m_{\Phi}^2, m_{\Phi}^2] - \frac{3}{16\pi^2} \lambda_7^I A_0[m_{\Phi}^2],
\end{align}
where the $A_0$ ($B_0$) function is the one-point (two-point) Passarino--Veltmann function, whose definition is given in appendix~\ref{sec:Loopfunc}.
As a result, we obtain
\begin{align}
    \widehat{\Pi}_{1 2}^{\lambda_7} (p^2) &= \frac{3}{16 \pi^2} \lambda_3 \lambda_7^R v^2 \big( B_0 [p^2; m_{\Phi}^2, m_{\Phi}^2] - B_0 [0; m_{\Phi}^2, m_{\Phi}^2] \big), \notag \\
    \widehat{\Pi}_{1 3}^{\lambda_7} (p^2) &= -\frac{3}{16 \pi^2} \lambda_3 \lambda_7^I v^2 \big( B_0 [p^2; m_{\Phi}^2, m_{\Phi}^2] - B_0 [0; m_{\Phi}^2, m_{\Phi}^2] \big),
\end{align}
where we have used $\Gamma_{H_2}/v \supset 3 \lambda_7^R A_0[m_{\Phi}^2] / 16\pi^2$ and $\Gamma_{H_3}/v \supset -3 \lambda_7^I A_0[m_{\Phi}^2] / 16\pi^2$ \footnote{
We have neglected the tadpole diagrams not including $\lambda_7$ because such contributions do not cause the EDMs in our scenario.
}.
The total $\widehat{\Pi}_{12}^{\lambda_7}(p^2)$ and $\widehat{\Pi}_{13}^{\lambda_7}(p^2)$ are UV finite, because 
\begin{align}
    B_0 [p^2; m_{\Phi}^2, m_{\Phi}^2] - B_0 [0; m_{\Phi}^2, m_{\Phi}^2] = - \int_0^1 dx~ \log \Big[ \frac{p^2 x^2 - p^2 x + m_{\Phi}^2}{m_{\Phi}^2} \Big].
\end{align}

By using these one-loop renormalized self energies, we obtain the top-quark CEDM at the two-loop level as
\begin{align}
    &\tilde{d}_t^{(2)} = \frac{\mathrm{Im}[\lambda_7\rho_{tt}]}{\sqrt{2}} \frac{3 \lambda_3 v}{(16 \pi^2)^2} \frac{2 m_t^2}{m_\Phi^2 - m_{H_1}^2}~\times \notag \\
    &\int_{4m_{\Phi}^2}^\infty ds~ \frac{\lambda^{1/2}(s,m_\Phi^2, m_\Phi^2)}{s} \Bigg\{ \frac{1}{m_{\Phi}^2 - s} \Big( C_{11}[s,t,t] - C_{11}[\Phi,t,t] \Big) \notag \\
    &- \frac{1}{m_{H_1}^2 - s} \Big( C_{11}[s,t,t] - C_{11}[H_1,t,t] \Big) - \frac{1}{s} \Big( C_{11}[\Phi,t,t] - C_{11}[H_1,t,t] \Big) \Bigg\},
    \label{eq:topCEDM2}
\end{align}
where $\lambda(a,b,c) = (a - b - c)^2 - 4 bc$.
To reduce the Feynman integral in the two-loop formula, we have used the dispersive approach~\cite{Bauberger:1994hx,Hollik:2005va,Aleksejevs:2018tfr}, in which the finite part of the $B_0$ function is rewritten by 
\begin{align}
    B (l^2; m_a^2, m_b^2) = \frac{1}{\pi} \int_{(m_a + m_b)^2}^\infty ds~ \frac{\mathrm{Im}\big[B[s; m_a^2,m_b^2] \big]}{s- l^2 - i\varepsilon} = \int_{(m_a + m_b)^2}^\infty ds~ \frac{\lambda^{1/2} (s,m_a^2, m_b^2)}{s(s- l^2 - i\varepsilon)}.
\end{align}
We note that the expression of $\tilde{d}_t^{(2)}$ depends on the renormalization scheme.
However, the scheme conversion of the input parameters compensates for such a difference, and we obtain the same prediction at the two-loop level in different renormalization schemes.
We will discuss this issue in appendix~\ref{sec:Renorm}.

In figure~\ref{fig:nEDM_1loop}, we show a prediction for $|d_n|$ obtained by the two-loop contributions as the single red point. 
We have taken $\alpha_1 =0$ and $m_{H_2} = 350$~GeV following our scenario. The other parameters are taken as $\lambda_7 = 1e^{i \pi/4}$ and $M = 30$~GeV. 
The size of the neutron EDM induced by the two-loop contributions is $|d_n|=2.2 \times 10^{-29}~e~\mathrm{cm}$. 

\subsection{Electron EDM \label{sec:eEDM}}

The top-quark EDM induces the electron EDM~\cite{Cirigliano:2016njn, Cirigliano:2016nyn,Fuyuto:2017xup}.
According to ref.~\cite{Fuyuto:2017xup}, by integrating out the additional scalar bosons at a scale $\Lambda$ above the EW scale, we get the effective operators
\begin{align}
    \mathcal{L}_{\mathrm{eff}} = -\frac{1}{\Lambda^2} \Big( \frac{g^\prime}{\sqrt{2}} C_{tB} \overline{Q_L} \sigma^{\mu \nu} t_R \tilde{\Phi}_1 B_{\mu \nu} +\frac{g}{\sqrt{2}} C_{tW} \overline{Q_L} \sigma^{\mu \nu} t_R \tau^a \tilde{\Phi}_1 W^a_{\mu \nu} + \mathrm{h.c.} \Big),
\end{align}
where $B_{\mu \nu}$ and $W_{\mu \nu}^a$ are the field strength tensors of the $\mathrm{U}(1)_Y$ and $\mathrm{SU}(2)_L$ gauge fields, respectively.
After the EW symmetry breaking, we obtain the EDM operators for $B_\mu$ and $W^a_\mu$ as
\begin{align}
    d_t^{B_\mu} = \frac{g_1 v}{\Lambda^2} \mathrm{Im}[C_{tB}], ~~
    d_t^{W^3_\mu} = \frac{g_2 v}{\Lambda^2} \mathrm{Im}[C_{tW}].
\end{align}
Then, after the renormalization group flow, by integrating out the top quark and the Z and W bosons at a scale $\mu$ below $\Lambda$, the electron EDM is obtained as~\cite{Fuyuto:2017xup}
\begin{align}
    d_e = -\frac{e}{2v} \Big( \frac{v}{\Lambda} \Big)^2 \Big(\log \frac{\Lambda}{\mu} \Big)^2 \bigg[ (A_e - D_e) \mathrm{Im}[C_{tB}] +(B_e - E_e) \mathrm{Im}[C_{tW}] \bigg], 
\end{align}
where
\begin{alignat}{3}
    &A_e = \mathcal{Y}_e ( 15 g_1^2 + 3g_2^2 ), &~~ &B_e = 10 \mathcal{Y}_e g_2^2, \notag \\
    &D_e = -6 \mathcal{Y}_e g_1^2, &~~ &E_e = -5 \mathcal{Y}_e (g_1^2 + g_2^2 ),
\end{alignat}
with $\mathcal{Y}_e = N_c y_e y_t /(4 \pi)^4$. 
In our calculation, we take the matching scale as $\Lambda = m_{\Phi}$.
The scale $\mu$ should be set around the EW scale.
In ref.~\cite{Fuyuto:2017xup}, $\mu = v$ is taken assuming $v \ll \Lambda$.
In our scenario, $\Lambda$ is rather close to $v$, and the prediction is sensitive to the choice of the scale $\mu$. In the numerical study, we study three cases $\mu = v, m_t$ and $m_Z$ and discuss the $\mu$ dependence on the electron EDM calculation.

Similar to the top-quark CEDM, the top-quark EDM vanishes at the one-loop level in our scenario.
In addition, due to the absence of $\rho_{ee}$, the Barr--Zee diagrams~\cite{Barr:1990vd} do not directly contribute to the electron EDM.
Therefore, the top-quark EDM at the two-loop level gives the leading contributions for the electron EDM.

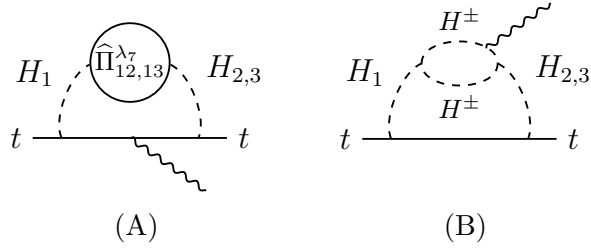
\begin{figure}[t]
    \centering
    \setlength{\feynhandlinesize}{0.7pt}
    \setlength{\feynhandarrowsize}{4pt}
    \begin{minipage}[t]{0.25\columnwidth}
    \centering
    \begin{tikzpicture}
        \begin{feynhand}
        \vertex (a) at (0,0){$t$}; \vertex[particle] (b) at (3,0){$t$};
        \vertex (c) at (0.6,0); \vertex (d) at (2.4,0);
        \vertex [ringblob] (g) at (1.5,1.0) {\footnotesize{$\widehat{\Pi}_{12,13}^{\lambda_7}$}};
        \vertex (e) at (1.5,0);\vertex (f) at (2.5,-0.7);
        \propag[plain] (a) to (b);
        \propag[sca] (c) to [quarter left, looseness=0.7, edge label=$H_1$] (g);
        \propag[sca] (g) to [quarter left, looseness=0.7, edge label=$H_{2,3}$] (d);
        \propag[pho] (f) to (e);
        \end{feynhand}
    \end{tikzpicture}
    \subcaption*{(A)}
  \end{minipage}
  ~~
  \begin{minipage}[t]{0.25\columnwidth}
    \centering
    \begin{tikzpicture}[baseline=-7mm]
        \begin{feynhand}
        \vertex (a) at (0,0){$t$}; \vertex[particle] (b) at (3,0){$t$};
        \vertex (c) at (0.6,0); \vertex (d) at (2.4,0);
        \vertex (g) at (1.0,1.0);
        \vertex (h) at (2.0,1.0);
        \vertex (e) at (1.85,1.2);
        \vertex (f) at (2.7, 1.8);
        \propag[plain] (a) to (b);
        \propag[sca] (c) to [quarter left, looseness=0.7, edge label=$H_1$] (g);
        \propag[sca] (h) to [quarter left, looseness=0.7, edge label=$H_{2,3}$] (d);
        \propag[pho] (f) to (e);
        \propag[sca] (h) to [half left, edge label=\footnotesize{$H^\pm$}] (g);
        \propag[sca] (g) to [half left, edge label=\footnotesize{$H^\pm$}] (h);
        \end{feynhand}
    \end{tikzpicture}
    \subcaption*{(B)}
  \end{minipage}
  \caption{Two-loop $\mathcal{O}(\lambda_7 \rho_{tt})$ diagrams for the top-quark EDM.}
  \label{fig:EDMdiagram1}
\end{figure}
In figure~\ref{fig:EDMdiagram1}, up to $\mathcal{O}(\lambda_7 \rho_{tt})$, the two-loop diagrams are shown.
The EDM is given by 
\begin{align}
    d_t^{(2)} = d_t^{(2), A} + d_t^{(2), B}, 
\end{align}
where each term on the right-hand side is expressed as
\begin{align}
    &d_t^{(2),A} = eQ_t \frac{\mathrm{Im}[\lambda_7 \rho_{tt}]}{\sqrt{2}} \frac{3 \lambda_3 v}{(16 \pi^2)^2} \frac{2 m_t^2}{m_\Phi^2 - m_{H_1}^2}~\times \notag \\
    &\int_{4m_{\Phi}^2}^\infty ds~ \frac{\lambda^{1/2}(s,m_\Phi^2, m_\Phi^2)}{s} \Bigg\{ \frac{1}{m_{\Phi}^2 - s} \Big( C_{11}[s,t,t] - C_{11}[\Phi,t,t] \Big) \notag \\
    &- \frac{1}{m_{H_1}^2 - s} \Big( C_{11}[s,t,t] - C_{11}[H_1,t,t] \Big) - \frac{1}{s} \Big( C_{11}[\Phi,t,t] - C_{11}[H_1,t,t] \Big) \Bigg\}, \notag \\
    &d_t^{(2),B} = - 2e \frac{\mathrm{Im}[\lambda_7 \rho_{tt}]}{\sqrt{2}} \frac{\lambda_3 v}{(16\pi^2)^2}~ \times \notag \\
    &\int_{4 m_\Phi^2}^\infty ds~ \frac{\lambda^{1/2}(s;m_{\Phi}^2,m_{\Phi}^2)}{s(m_\Phi^2-s)(m_{H_1}^2 - s)} \Bigg[ m_{H_1}^2 C_0[t,\Phi,H_1] + s C_0 [t,s,s] - s C_0[t, \Phi, s] \notag \\
    &- m_{H_1}^2 C_0[t,s,H_1] + B_0[0;s,m_\Phi^2] - B_0[0;s,s] + B_0[0;m_{H_1}^2,s] - B_0[0;m_{H_1}^2,m_\Phi^2] \notag \\
    &-2m_t^2 \Big\{ C_{23}[H_1,\Phi,t] - C_{23}[\Phi,H_1,t] -C_{23}[s,\Phi,t] + C_{23}[\Phi,s,t] \notag \\
    &-C_{23}[H_1,s,t] + C_{23}[s,H_1,t] \Big\} \Bigg].
    \label{eq:topEDM1}
\end{align}
Again, the definition of the loop functions is given in appendix~\ref{sec:Loopfunc}, and $Q_t = 2/3$ is the electromagnetic charge for the top quarks.

The corresponding EDMs for $B_\mu$ and $W^3_\mu$ are obtained by replacing charges as
\begin{align}
    &d_t^{B_\mu} = d_t^{(2),A} \Big|_{eQ_t \to 5g_1/12} + d_t^{(2),B} \Big|_{e \to g_1/2}, \notag \\
    &d_t^{W^3_\mu} = d_t^{(2),A} \Big|_{eQ_t \to g_2/4} + d_t^{(2),B} \Big|_{e \to g_2/2}.
\end{align}

\section{Results \label{sec:BAU_and_EDM}}

\begin{table}[t]
    \centering
    \scalebox{0.87}[0.87]{
    \begin{tabular}{|l|l|}
    \hline
    EDMs  & Current bounds ($90\%$ C.L.) \\ \hline \hline
    $d_e$ (ACME-II) & $1.1 \times 10^{-29}$ $e$ cm ~\cite{ACME:2018yjb}  \\ \hline 
    $d_e$ (JILA) &  $4.1 \times 10^{-30}$ $e$ cm ~\cite{Roussy:2022cmp} \\ \hline 
    $d_n$ (nEDM)  & $1.8 \times 10^{-26}$ $e$ cm ~\cite{Abel:2020pzs}  \\ \hline 
    $d_p$ (U.Wash.) & $2.1 \times 10^{-25}$ $e$ cm ~\cite{Graner:2016ses,Sahoo:2016zvr} \\ \hline
    \end{tabular}}
    \scalebox{0.87}[0.87]{
    \begin{tabular}{|l|l|}
        \hline
        EDMs  & Future prospect  \\ \hline \hline
        $d_e$ (ACME-III) & $3 \times 10^{-31}$ $e$ cm~\cite{Hiramoto:2022fyg} \\ \hline 
        $d_e$ ($\mathrm{EDM}^3$) & $\mathcal{O}(10^{-33})$ $e$ cm~\cite{Vutha:2018tsz,Ardu:2024bxg} \\ \hline 
        $d_n$ (nEDM)  & $2 \times 10^{-28}$ $e$ cm~\cite{nEDM:2019qgk} \\ \hline 
        $d_p$ (Storage Ring) & $\mathcal{O}(10^{-29})$ $e$ cm~\cite{Alarcon:2022ero} \\ \hline
        \end{tabular}}
        \caption{Current limits and future sensitivities of the EDMs.}
    \label{tab:EDMbounds}
\end{table}

In this section, we show numerical results for the BAU and EDMs in our scenario.
We scan the following parameter region.
\begin{alignat}{3}
    m_{\Phi} &= [200,500]~\mathrm{GeV}\qc&
    M &= [0,m_{\Phi}]~\mathrm{GeV}\qc&
    \lambda_2 &= [0,1],\notag\\
    |\lambda_7| &= [0,1]\qc& |\rho_{tt}|&=[0,0.5]\qc& \mathrm{arg}[\lambda_7 \rho_{tt}] &= -\pi/2.
\end{alignat}
Our parameter scan is subject to the theoretical constraints such as the perturbative unitarity~\cite{Kanemura:1993hm,Akeroyd:2000wc,Ginzburg:2005dt,Kanemura:2015ska} and the bounded from below conditions~\cite{Deshpande:1977rw,Klimenko:1984qx,Sher:1988mj,Nie:1998yn,Kanemura:1999xf,Kanemura:2000bq,Ferreira:2004yd,Bahl:2022lio}.
The lower bound on $m_{\Phi}$ and the upper bound on $|\rho_{tt}|$ come from the direct search for the $H^\pm \to t b$ decay mode~\cite{ATLAS:2021upq}.
The tree-level $H_{1}$ couplings are SM-like in our scenario because we take $\lambda_{6}=0$, and there is no mixing\footnote{If $\lambda_{6}\neq0$, there are mixing among the neutral scalars. Since the $H_{1}$ couplings are changed from their SM values at the tree level, the size of mixing angles is constrained by the Higgs signal measurements  (see the discussion in section~\ref{sec:EWBG}). In addition, the additional neutral scalars decay into $H_{1}$, e.g. $H_{2}\to H_{1}H_{1}$ and $H_{3}\to ZH_{1}$ if they are kinematically accessible. Direct searches of these channels also provide strong constraints on the size of mixing angles (See ref.~\cite{Aiko:2020ksl, Karan:2023kyj} for example).}.
However, the loop-induced $H_1 \gamma \gamma$ coupling can deviate from the SM value through the charged Higgs contributions~\cite{Ellis:1975ap,Shifman:1979eb, Gavela:1981ri, Barroso:1999bf, Arhrib:2003vip, Djouadi:2005gj, Akeroyd:2007yh, Posch:2010hx, Kanemura:2016sos, Degrassi:2023eii, Aiko:2023nqj}.
In the above scan region, the largest deviation in the $H_1 \gamma \gamma$ coupling is achieved for $m_{H^\pm}=500~\mathrm{GeV}$ and $M=0$.
At this point, we obtain $\mu^{\gamma \gamma} = (\sigma \mathrm{Br}_{\gamma \gamma})_{\mathrm{2HDM}} / (\sigma \mathrm{Br}_{\gamma \gamma})_{\mathrm{SM}} \simeq \Gamma^{\gamma \gamma}_{\mathrm{2HDM}} / \Gamma^{\gamma \gamma}_{\mathrm{SM}}= 0.9$ at the one-loop level, where $\Gamma^{\gamma \gamma}$ is the decay rate for $H_1 \to \gamma \gamma$.
This is consistent with the current result $\mu_{\mathrm{exp}}^{\gamma \gamma} = 1.04^{+0.10}_{-0.09}$~\cite{ATLAS:2022tnm} at the $2 \sigma$ level.

The generated BAU $\eta_B = (n_B - n_{\overline{B}})/s$ is calculated by solving the transport equations, which are explicitly given in refs.~\cite{Fromme:2006cm,Cline:2020jre,Enomoto:2022rrl}.
The calculations are parallel to those studies, but we have applied the interaction rates in the collision term summarized in ref.~\cite{Cline:2021dkf}.
In addition to the model parameters, we scan the wall velocity as
\begin{align}
    v_w = [0.1,1/\sqrt{3}].
\end{align}
\begin{figure}[t]
    \centering
    \includegraphics[width=1.0\linewidth]{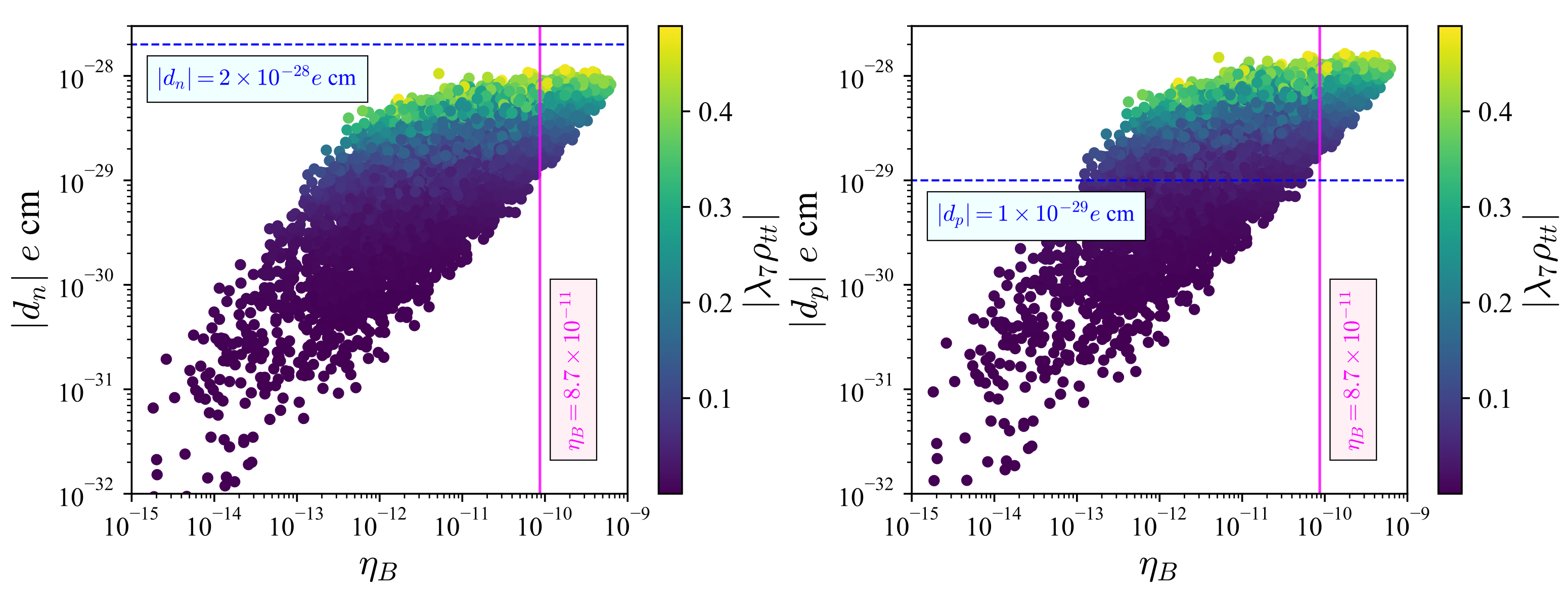}
    \caption{Left (right): Correlation between the neutron (proton) EDM and the generated baryon asymmetry.
    The input parameters are scanned as $m_{\Phi} = [200,500]~\mathrm{GeV}$, $M=[0,m_{\Phi}]$, $\lambda_2 = [0,1]$, $|\lambda_7| = [0,1]$, $|\rho_{tt}|=[0,0.5]$, $\mathrm{arg}[\lambda_7 \rho_{tt}] = -\pi/2$ and $v_w = [0.1,1/\sqrt{3}]$.
    The colored points show $|\lambda_7 \rho_{tt}|$.
    The vertical magenta solid line corresponds to the observed BAU.
    The horizontal blue dashed line in each panel is a future sensitivity for the corresponding EDM.}
    \label{fig:Corr_BAU_npEDM}
\end{figure}
\begin{figure}[p]
    \centering
    \includegraphics[width=0.56\linewidth]{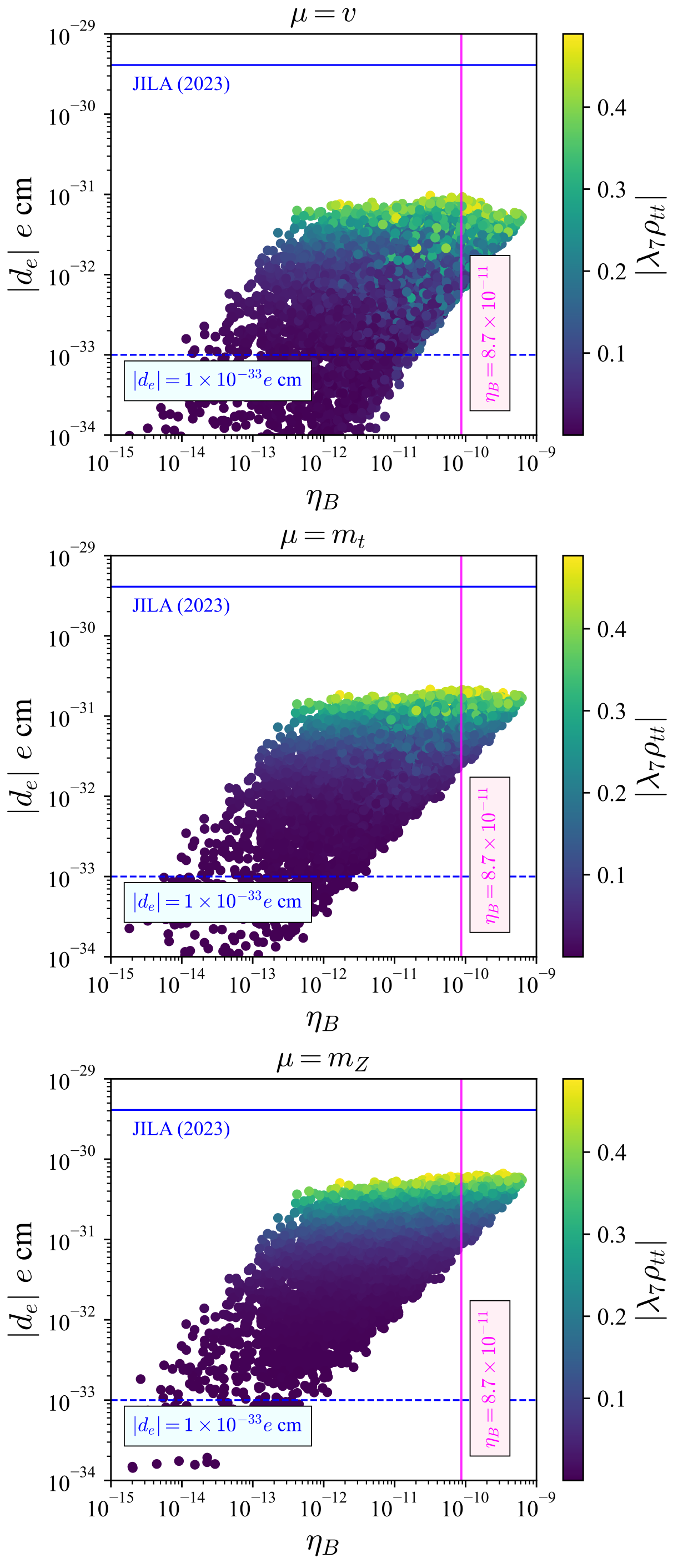}
    \caption{Correlations between the electron EDM and the generated baryon asymmetry.
    From above to below, $\mu = v,m_t,m_Z$ are taken, respectively.
    The horizontal blue solid lines are the current bound from JILA~\cite{Roussy:2022cmp}, and the dashed lines are a future sensitivity~\cite{Vutha:2018tsz,Ardu:2024bxg}.
    }
    \label{fig:Corr_BAU_eEDMs}
\end{figure}
In the left (right) panel of figure~\ref{fig:Corr_BAU_npEDM}, we show the correlation between the neutron (proton) EDM and the BAU.
The color gradient indicates the value of $|\lambda_7 \rho_{tt}|$.
The observed BAU $\eta_{B} \simeq 8.7 \times 10^{-11}$~\cite{Planck:2018vyg} is shown by the pink solid lines.
The current limits for the neutron and proton EDMs are $|d_n| = 1.8 \times 10^{-26}~e~\mathrm{cm}$ by nEDM~\cite{Abel:2020pzs} and $|d_p| = 2.1 \times 10^{-25}~e~\mathrm{cm}$ by U. Wash.~\cite{Graner:2016ses,Sahoo:2016zvr}, respectively.
These limits are outside of the figures.
The future prospects are $|d_n| = 2 \times 10^{-28}~e~\mathrm{cm}$~\cite{nEDM:2019qgk} and $|d_p| = \mathcal{O}(10^{-29})~e~\mathrm{cm}$~\cite{Alarcon:2022ero}, respectively.
They are shown by the blue dashed lines.
In table~\ref{tab:EDMbounds}, the current status of several EDM bounds and the future prospects are summarized.

From the left panel of figure~\ref{fig:Corr_BAU_npEDM}, we can see that the neutron EDM grows as $|\lambda_7 \rho_{tt}|$ becomes larger. On the other hand, the predicted BAU does not show such simple behavior. This is because it much depends on the other parameters, $m_{\Phi}$, $M$, $\lambda_2$ and $v_w$. The predictions on the proton EDM show almost the same behavior as that of the neutron EDM (see the right panel of figure~\ref{fig:Corr_BAU_npEDM}).
To realize the observed BAU, the size of the neutron and proton EDMs should be larger than $\mathcal{O}(10^{-29}) ~e~\mathrm{cm}$.
Our scenario is comparable with the current limits~\cite{Abel:2020pzs,Graner:2016ses,Sahoo:2016zvr}. 
Also, it can be tested at the future experiments~\cite{Alarcon:2022ero} whose sensitivities are expected to be $\mathcal{O}(10^{-29})~e~\mathrm{cm}$.

In figure~\ref{fig:Corr_BAU_eEDMs}, we show the correlations between the electron EDM and the BAU.
As discussed in section~\ref{sec:eEDM}, the prediction for the electron EDM depends on $\log (m_{\Phi}/\mu)$.
In figure~\ref{fig:Corr_BAU_eEDMs}, we have taken $\mu = v,m_t,m_Z$ from above to below.
The solid blue lines show the current bound on the electron EDM by JILA~\cite{Roussy:2022cmp}, $|d_e| = 4.1 \times 10^{-30}~e~\mathrm{cm}$.
The blue dashed lines show the expected future sensitivity, $|d_e|=\mathcal{O}(10^{-33})~e~\mathrm{cm}$~\cite{Vutha:2018tsz,Ardu:2024bxg}.
Similar to the neutron and proton EDMs, the electron EDM should be larger than $4 \times 10^{-33}~e~\mathrm{cm}$ (for $\mu=v$) or more to realize the observed BAU, though the lowest value depends on the choice of $\mu$.
In all three cases, our scenario can be tested in future experiments~\cite{Vutha:2018tsz,Ardu:2024bxg}.

\begin{figure}[t]
    \centering
    \includegraphics[width=1\linewidth]{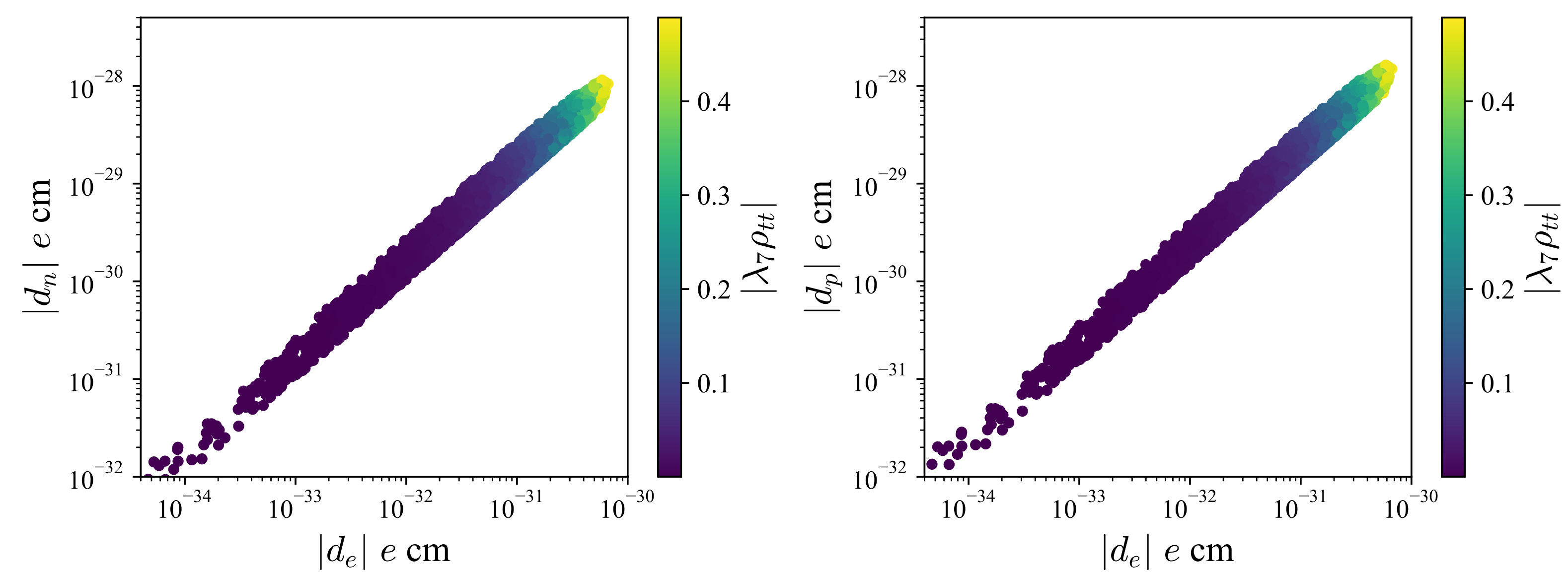}
    \caption{Left (right): Correlation between the neutron (proton) EDM and the electron EDM.
    Here, $\mu = m_Z$ is taken.}
    \label{fig:Corr_np_and_e}
\end{figure}
Finally, let us show a correlation between the electron EDM and the neutron (proton) EDM. The results are shown in figure~\ref{fig:Corr_np_and_e}.
Here, we take $\mu = m_Z$.
If we take $\mu = v$ instead, the predicted values of the electron EDM are about 10 times smaller than the values shown in the figure.
It is found that there is a strong correlation between the electron EDM and the neutron (proton) EDM.
This is because both are induced by the CPV parameter $\mathrm{Im}[\lambda_7 \rho_{tt}]$.
This strong correlation is a characteristic prediction in our scenario.
Thus, our scenario can be verified by studying the correlations among various EDMs.

\section{Discussion \label{sec:discussion}}

In this section, we give discussions in order.

Let us first discuss how to test our scenario in other experiments.
The first-order EW phase transition requires sizable loop corrections of the additional scalar bosons. 
Such effects could be probed by measuring the triple Higgs coupling and the Higgs di-photon decay rate~\cite{Kanemura:2004ch,Ellis:1975ap,Shifman:1979eb, Gavela:1981ri, Barroso:1999bf, Arhrib:2003vip, Djouadi:2005gj, Akeroyd:2007yh, Posch:2010hx, Kanemura:2016sos, Degrassi:2023eii, Aiko:2023nqj}.
In addition, our scenario requires non-zero $\rho_{tt}$.
It can be probed by direct searches of $H^\pm \to tb$ and $H_{2,3} \to t \overline{t}$ if the processes are kinematically allowed.
Future colliders, such as HL-LHC, may be useful to discover signatures of our model and improve sensitivities to $\rho_{tt}$.
The CPV effects could also be tested at collider experiments.
They might appear in the decay of the additional scalar bosons into a pair of top quarks~\cite{Bernreuther:1993hq,Bernreuther:1998qv,Khater:2003wq,Hagiwara:2013jp}.
In ref.~\cite{Kanemura:2024ezz}, it was pointed out that $H^\pm\to W^{\pm}Z$ proceeding via radiative corrections~\cite{Grifols:1980uq,CapdequiPeyranere:1990qk,Kanemura:1997ej,Diaz-Cruz:2001thx,Arhrib:2006wd,Abbas:2018pfp,Aiko:2021can} is sensitive to $\mathrm{Im}[\lambda_7 \rho_{tt}]$, when there is a mass difference between the charged scalar boson and one of the additional neutral scalar bosons.
The stochastic gravitational waves produced by the first-order phase transition would be measured by future space-based gravitational waves observables~\cite{Grojean:2006bp,Espinosa:2010hh,Caprini:2015zlo, Kakizaki:2015wua, Hashino:2016rvx}.

Second, we have shown that the BAU is dominantly generated by $\mathrm{Im}[\lambda_7 \rho_{tt}]$, while the other CP phases are less effective. 
Hence, we have assumed that the interactions of the additional scalar bosons irrelevant to EWBG are zero.
However, this assumption is not stable under radiative corrections because they are not protected by symmetries\footnote{We note that $(\rho^f)_{lm}$ would not be changed by renormalization group evolutions below the mass scale of the additional scalar bosons.}.
Let us consider the case when the other couplings are additionally introduced. 
When the Yukawa coupling, $(\rho^f)_{lm}$, other than those of the top quark are turned on, although the generated baryon asymmetry would not be changed drastically from our scenario, the EDMs receive extra contributions. 
Then, thanks to such additional degrees of freedom, it might be possible to realize EDM cancellations by adjusting those CPV parameters (cf.~\cite{Bian:2014zka, Cheung:2020ugr, Kanemura:2020ibp}).

Third, if we additionally turn on $\mathrm{Im}[\lambda_5 \rho_{tt}^2]$, although the predicted BAU is not changed drastically, the top-quark (C)EDM appear at the one-loop level (see the right panel of figure~\ref{fig:CEDMdiagram1}).
Also, it might be possible that $\lambda_5$ is non-zero with keeping $\mathrm{Im}[\lambda_5 \rho_{tt}^2] = 0$, i.e., $\mathrm{arg}[\lambda_5]=-2\mathrm{arg}[\rho_{tt}]$.
Then, the top-quark (C)EDM appears at the two-loop level, and there are two types of contributions involving $\mathrm{Im}[\lambda_5^* \lambda_7^2]$ or $\mathrm{Im}[\lambda_7 \rho_{tt}]$.
As long as the size of $\lambda_5$ is small, the former would become sub-leading.

Fourth, we have treated the wall velocity as a free parameter in our numerical analysis of the BAU.
Not only the BAU but also the gravitational waves produced by the first-order EW phase transition~\cite{Grojean:2006bp,Espinosa:2010hh,Caprini:2015zlo, Kakizaki:2015wua, Hashino:2016rvx} depend on the bubble wall velocity, and therefore, it is important to determine its value from first principles.
The wall velocity has been evaluated in refs.~\cite{Moore:1995ua, Moore:1995si,Konstandin:2014zta, Kozaczuk:2015owa, Dorsch:2016nrg, Cline:2021iff,Jiang:2022btc} in several models when their particle contents are similar to the SM.
The wall velocity is estimated to be $\mathcal{O}(10^{-1})$.
See refs.~\cite{Laurent:2022jrs,Ai:2023see,Ekstedt:2024fyq} for recent discussions.

Finally, in this paper, we have used the semi-classical approach~\cite{Cline:2000nw,Fromme:2006wx,Joyce:1994fu,Joyce:1994zn,Joyce:1994zt,Cline:2020jre} to evaluate the BAU.
Alternatively, if the VEV insertion approximation~\cite{Riotto:1995hh,Riotto:1997vy} is adopted for deriving the transport equations, it had been argued that the produced BAU is enhanced by several orders of magnitude compared to the semi-classical approach~\cite{Basler:2021kgq,Cline:2021dkf}.
However, the CPV sources may disappear at the leading order of the gradient expansion in the approach with the VEV insertion approximation~\cite{Kainulainen:2021oqs,Postma:2022dbr}.
On the other hand, there are discussions based on other approaches, such as the VEV resummation scheme~\cite{Cirigliano:2009yt,Cirigliano:2011di,Li:2024mts}.
Such frameworks may give larger BAU than that in the semi-classical approach, though detailed analysis is beyond the scope of this paper.

\section{Conclusions \label{sec:conclusion}}

In this paper, we have studied two Higgs doublet models with successful electroweak baryogenesis but without EDM cancellations. 
The additional scalar bosons are favored to couple mainly with the top quark with CP violations for EWBG to work sufficiently.
The CP violations relevant to the BAU generally contribute to the EDMs simultaneously. 
Instead of introducing extra CP phases irrelevant to the BAU for satisfying the EDM constraints, we have considered a scenario where the light-fermion couplings are suppressed. 
It has been found that the leading contributions arise in the top-quark EDMs at the two-loop level, leading to the electron, neutron, and proton EDMs via radiative corrections.
Since there are no additional CP phases, they are correlated with the baryon asymmetry.
We have shown that our scenario is compatible with the current experimental bounds and is within the scope of future EDM experiments.

\section*{Acknowledgment}
This work is supported by the JSPS Grant-in-Aid for JSPS Fellows No.~22KJ3126 [M.A.], No.~23KJ1460 [Y.M.].
This work is also supported by JSPS KAKENHI Grant Numbers 21H01086 [M.E.], 22K21347 [M.E.], 23K20847 [M.E.], 20H00160 [S.K.] and 23K17691 [S.K.].
All Feynman diagrams are drawn with \texttt{TikZ-FeynHand}~\cite{Ellis:2016jkw,Dohse:2018vqo}.

\appendix
\section{Loop functions \label{sec:Loopfunc}}

We define the loop functions used in the EDM calculations.
The Passarino--Veltmann functions~\cite{tHooft:1978jhc,Passarino:1978jh} are given by
\begin{align}
    &\frac{1}{16\pi^2 } B_0 [k; m_a^2, m_b^2] = \mu^{2\epsilon} \int \frac{d^D l}{i (2\pi)^D} \frac{1}{l^2 - m_a^2 + i\varepsilon}\frac{1}{(l+k)^2 - m_b^2 + i\varepsilon},\notag \\
    &\frac{1}{16\pi^2 } C_{\{ 0,\mu,\mu \nu \}} [k_1, k_2; m_a^2,m_b^2,m_c^2] =\notag \\
    &\qquad \mu^{2\epsilon} \int \frac{d^D l}{i (2\pi)^D} \frac{\{1, l_\mu, l_\mu l_\nu\}}{l^2 - m_a^2 + i\varepsilon}\frac{1}{(l+k_1)^2 - m_b^2 + i\varepsilon} \frac{1}{(l+k_1 + k_2)^2 - m_c^2 + i\varepsilon},
\end{align}
where $\mu$ is the mass dimensional parameter and $D=4-2 \epsilon$ is the space-time dimension.
The tensor decompositions are given by
\begin{align}
  &C_{\mu}[k_1, k_2; m_a^2,m_b^2,m_c^2] = C_{11} k_{1 \mu} + C_{12} k_{2 \mu}, \notag \\
  &C_{\mu \nu}[k_1, k_2; m_a^2,m_b^2,m_c^2] = C_{21} k_{1 \mu}k_{1 \nu} + C_{22} k_{2 \mu} k_{2 \nu} + C_{23} (k_{1\mu} k_{2 \nu} + k_{2 \mu} k_{1 \nu}) + C_{24} g_{\mu\nu}.
\end{align}
Especially, the $C$ type functions used in section~\ref{sec:EDM} are given by 
\begin{align}
    &C_{11} [a, t, t] = \int_0^1 dx \frac{x^2}{m_t^2 x^2 - m_a^2 x + m_a^2}, \notag \\
    &C_0[t,a,b] =
    \begin{cases}
        -\int_0^1 dx~ \frac{1}{c} \Big( \log\frac{D}{D-cx} \Big) ,~~(m_a \neq m_b) \notag \\
        - \int_0^1 dx~\frac{x}{D},~~(m_a = m_b)
    \end{cases} \notag \\
    &C_{23}[a, b, t] = 
    \begin{cases}
        \int_0^1 dx~ \frac{x-1}{c^2} \Big( \big( D+c(x-1) \big) \log \frac{D}{D + c x} + c x \Big) ,~~(m_a \neq m_b) \notag \\
        \int_0^1 dx~\frac{1}{D} \Big( -\frac{1}{2}x^3 + \frac{3}{2}x^2 - x \Big),~~(m_a = m_b)
    \end{cases}
\end{align}
where 
\begin{align}
    D = m_t^2 x^2 + (m_a^2 - 2 m_t^2) x + m_t^2, ~~~c = m_b^2 -m_a^2.
\end{align}

\section{Other renormalization schemes \label{sec:Renorm}}

In this appendix, we discuss the renormalization scheme and scale dependencies in the calculations of the top-quark (C)EDM.
Since the discussion for the top-quark EDM is parallel to that of the CEDM, we discuss only the top-quark CEDM.
We perform the calculations of the top-quark CEDM in the $\MSbar$ and OS renormalization schemes in our scenario.
We show that the expression for $\tilde{d}_t^{(2)}$ is changed depending on the renormalization of the mixed self energies (see figure~\ref{fig:CEDMdiagram3}).
It is found that the scheme conversion of the input parameters compensates for such a difference.
Thus, we obtain the same prediction at the two-loop level in different renormalization schemes.
In the $\MSbar$ scheme, we also show that there is no renormalization scale dependence in the top-quark CEDM up to the two-loop level.
In the following discussions, we take $\chi = -\mathrm{arg}[\lambda_5]/2$, and we only focus on the CPV contributions.

\subsection{EP scheme and \texorpdfstring{$\MSbar$}{MSbar} scheme}
In the $\MSbar$ scheme, the parts of the renormalized mixed self energies involving $\lambda_7$ are obtained as 
\begin{align}
    \widehat{\Pi}_{12}^{\lambda_{7},\, \MSbar}(p^{2})
    &=
    \eval{
    \frac{3}{16 \pi^2} \lambda_3 \lambda_7^R v^2 \big( B_0 [p^2; m_{\Phi}^2, m_{\Phi}^2] - B_0 [0; \mu^2, \mu^2] \big)
    }_{\MSbar},
    \notag \\
    \widehat{\Pi}_{1 3}^{\lambda_{7},\, \MSbar}(p^{2})
    &=
    \eval{
    -\frac{3}{16 \pi^2} \lambda_3 \lambda_7^I v^2 \big( B_0 [p^2; m_{\Phi}^2, m_{\Phi}^2] - B_0 [0; \mu^2, \mu^2] \big)
    }_{\MSbar},
    \label{eq: MSbar renormalized mixing self-eneries}
\end{align}
where $\mu$ is the mass dimensional parameter.
It should be understood that the right-hand sides of the above equations are evaluated with the $\overline{\mathrm{MS}}$ quantities.

Then, from figure~\ref{fig:CEDMdiagram3}, we obtain
\begin{align}
    \tilde{d}_t^{(2), \MSbar}
    &=
    \tilde{d}_t^{(2)}+\Delta \tilde{d}_t^{(2), \MSbar},
    \notag \\
    \Delta \tilde{d}_t^{(2), \MSbar}
    &=
    \eval{% \tilde{d}_t^{(2)}
    \frac{\mathrm{Im}[\lambda_7\rho_{tt}]}{\sqrt{2}} \frac{3 \lambda_3 v}{(16 \pi^2)^2} \frac{2m_t^2}{m_{\Phi}^2 - m_{H_1}^2}
    \Big( C_{11}[\Phi, t, t] - C_{11}[H_1, t, t] \Big) \log \frac{\mu^2}{m_{\Phi}^2}}_{\MSbar},
\end{align}
where $\tilde{d}_t^{(2)}$ is given in eq.~\eqref{eq:topCEDM2} while it is evaluated with the $\MSbar$ quantities not the EP quantities. 

To understand the origin of $\Delta \tilde{d}_t^{(2), \MSbar}$, we discuss the relation of the $\lambda_{6}$ coupling in the $\overline{\mathrm{MS}}$ and ${\rm EP}$ schemes.
From eq.~\eqref{eq: MSbar renormalized mixing self-eneries}, it is obtained as
\begin{align}
    \lambda_{6}^{\MSbar} =
    \eval{
    \lambda_6
    -\frac{3}{16 \pi^2} \lambda_3 \lambda_7 \log \frac{\mu^2}{m_\Phi^2}
    +\dots~
    }_{\text{EP}},
    \label{eq:l6MSl6EP}
\end{align}
where everything in the right-hand side is the EP quantities, and we have shown the EDM relevant parts.
In our scenario, we take $\lambda_6=0$, and there is no one-loop contribution in the EP scheme (see eq.~\eqref{eq:topCEDM1}).
On the other hand, in the $\MSbar$ scheme, we have the one-loop contributions.
\begin{align}
    \tilde{d}_t^{(1), \MSbar}
    &=
    \eval{
    \frac{\mathrm{Im}[\lambda_{6}  \rho_{tt}]}{\sqrt{2}} \frac{v}{16 \pi^2}   \frac{2m_t^2}{m_{\Phi}^2 - m_{H_1}^2} \Big( C_{11}[\Phi, t, t] - C_{11}[H_1, t, t] \Big)}_{\overline{\mathrm{MS}}},
    \label{eq: topCEDM1_MSbar}
\end{align}
where we have omitted $\mathcal{O}(\lambda_5 \rho_{tt}^2 )$ contributions because they only lead subleading $\mathcal{O}(\lambda_7^2 \rho_{tt}^2)$ contributions.
By substituting eq.~\eqref{eq:l6MSl6EP} into eq.~\eqref{eq: topCEDM1_MSbar}, we obtain
\begin{align}
    \tilde{d}_t^{(1),\MSbar}+ \tilde{d}_t^{(2),\MSbar}
    =\eval{\tilde{d}_t^{(2)}+\mathcal{O}(\hbar^{3})}_{\mathrm{EP}}.
\end{align}
Thus, the difference in the two-loop expressions corresponds to the scheme difference of the $\lambda_{6}$ coupling, and we obtain the same prediction with the EP scheme up to the two-loop order.
Since $\tilde{d}_t^{(2)}$ is $\mu$-independent, $\tilde{d}_t^{(1),\MSbar}+ \tilde{d}_t^{(2),\MSbar}$ is scale-independent up to the two-loop order.

\subsection{EP scheme and OS scheme}
In the OS scheme, we shift the bare masses of the scalar bosons as
\begin{align}
    m_{i, B}^{2} = m_{i}^{2}+\delta m_{i}^{2},
\end{align}
and the bare scalar fields $\phi_{i}$ are shifted as
\begin{align}
\phi_{i, B} &= \qty(1+\frac{\delta Z_{ii}}{2})\phi_{i}+\sum_{i\neq j}\frac{\delta Z_{ij}}{2}\phi_{j},
\label{eq: wf_scalar}
\end{align}
where $\phi_{i}$ represents a mass-eigen scalar field, i.e. $\phi_{i}=H_{1},H_{2},H_{3},H^{\pm},G^{0}$ or $G^{\pm}$.
For tadpole renormalization, we adopt the standard tadpole scheme. In this scheme, similarly to the EP scheme, we determine the tadpole counter terms so that the renormalized one-point functions vanish order by order (see eq.\eqref{eq:EP_onepoint}).
Then, the renormalized self-energies are obtained as
\begin{align}
\widehat{\Pi}_{ij}^{\mathrm{OS}}(p^{2}) &=
\Pi_{ij}(p^{2})
+\frac{\delta Z_{ij}}{2}\qty(p^{2}-m_{i}^{2})
+\frac{\delta Z_{ji}}{2}\qty(p^{2}-m_{j}^{2})
-\delta m_{i}^{2}\delta_{ij},
\label{eq: rSE_neutral}
\end{align}
where $\Pi_{ij}(p^{2})$ is a 1PI two-point function of scalar bosons.

We determine $\delta m_{i}^{2}$ so that the pole of the propagator coincides with $m_{i}^{2}$.
\begin{align}
\widehat{\Pi}_{ii}^{\mathrm{OS}}(m_{i}^{2}) = 0.
\end{align}
This condition leads
\begin{align}
    \delta m_{i}^{2} = \Pi_{ii}(m_{i}^{2}).
\end{align}
On the other hand, $\delta Z_{ii}$ are fixed so that the residue of the propagator becomes unity.
\begin{align}
\eval{\dv{p^{2}}\widehat{\Pi}_{ii}^{\mathrm{OS}}}_{p^{2}=m_{i}^{2}} = 0.
\end{align}
We obtain
\begin{align}
    \delta Z_{ii} = -\eval{\dv{p^{2}}\Pi_{ii}(p^{2})}_{p^{2}=m_{i}^{2}}.
\end{align}
The off-diagonal parts $\delta Z_{ij}\ (i\neq j)$ are determined so that the scalar mixings vanish at the pole position.
\begin{align}
\widehat{\Pi}_{ij}^{\mathrm{OS}}(m_{i}^{2})
=
\widehat{\Pi}_{ij}^{\mathrm{OS}}(m_{j}^{2})
=
0.
\end{align}
These conditions lead
\begin{align}
\frac{\delta Z_{ij}}{2} = \frac{\Pi_{ij}(m_{j}^{2})}{m_{i}^{2}-m_{j}^{2}}\qc
\frac{\delta Z_{ji}}{2} = -\frac{\Pi_{ij}(m_{i}^{2})}{m_{i}^{2}-m_{j}^{2}}.
\label{eq: delZ_ij}
\end{align}

\begin{figure}[t]
    \centering
    \setlength{\feynhandlinesize}{0.7pt}
    \setlength{\feynhandarrowsize}{5pt}
    ~~
    \begin{minipage}[t]{0.3\columnwidth}
    \centering
    \begin{tikzpicture}
        \begin{feynhand}
        \vertex (a) at (0,0){$t$}; \vertex[particle] (b) at (3,0){$t$};
        \vertex [crossdot] (c) at (0.6,0) {}; \vertex (d) at (2.4,0);
        \vertex (e) at (1.5,0);\vertex (f) at (2.5,-0.8);
        \propag[plain] (a) to (c);
        \propag[plain] (c) to (b);        
        \propag[sca] (c) to [half left, looseness=1.7] (d);
        \propag[glu] (e) to (f);
        \vertex (g) at (1.5,1.3) {$H_i$};
        \end{feynhand}
    \end{tikzpicture}
    \end{minipage}
    ~~
    \begin{minipage}[t]{0.3\columnwidth}
    \centering
    \begin{tikzpicture}
        \begin{feynhand}
        \vertex (a) at (0,0){$t$}; \vertex[particle] (b) at (3,0){$t$};
        \vertex (c) at (0.6,0); \vertex [crossdot] (d) at (2.4,0) {};
        \vertex (e) at (1.5,0);\vertex (f) at (2.5,-0.8);
        \propag[plain] (a) to (d);
        \propag[plain] (d) to (b);        
        \propag[sca] (c) to [half left, looseness=1.7] (d);
        \propag[glu] (e) to (f);
        \vertex (g) at (1.5,1.3) {$H_i$};
        \end{feynhand}
    \end{tikzpicture}
    \end{minipage}
    \caption{Two-loop diagrams of the top-quark CEDM with the counter terms of the top-quark Yukawa couplings.}
    \label{fig:CEDMdiagram_YCT}
\end{figure}
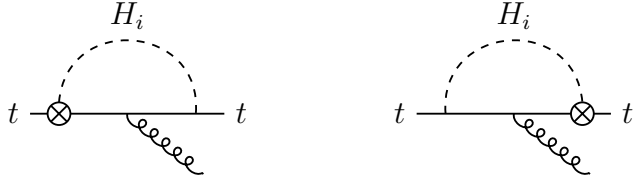

For the top-quark (C)EDM calculations, the counter terms of the top-quark Yukawa couplings are relevant (see figure~\ref{fig:CEDMdiagram_YCT}). 
Let us denote the renormalized $H_{i}tt$ couplings as
\begin{align}
    \widehat{\Gamma}_{H_{i}tt}^{L/R} = \Gamma_{H_{i}tt}^{L/R}+\delta\Gamma_{H_{i}tt}^{L/R},
\end{align}
where $\Gamma_{H_{i}tt}^{L/R}$ and $\delta\Gamma_{H_{i}tt}^{L/R}$ are the 1PI contributions and the counter terms, respectively.
The superscript $L$ ($R$) indicates the left (right) chiral part of the top-quark Yukawa couplings. 
In our scenario, CPV contributions originate from the mixing-angle and wave-function counter terms in $\delta\Gamma_{H_{i}tt}^{L/R}$.
We shift the mixing angles as
\begin{align}
\theta_{ij, B} = \theta_{ij}+\delta\theta_{ij}.
\end{align}
Then, the leading $\mathcal{O}(\lambda_{7} \rho_{tt})$ parts of $\delta \Gamma_{H_{i}tt}^{L/R}$ are obtained as
\begin{align}
\delta \Gamma_{H_{1}tt}^{L}
&=
g_{H_{2}tt}^{L}\qty(-\delta \theta_{12}+\frac{\delta Z_{21}}{2})
+g_{H_{3}tt}^{L}\qty(-\delta \theta_{13}+\frac{\delta Z_{31}}{2}), \\
\delta \Gamma_{H_{1}tt}^{R}
&=
g_{H_{2}tt}^{R}\qty(-\delta \theta_{12}+\frac{\delta Z_{21}}{2})
+g_{H_{3}tt}^{R}\qty(-\delta \theta_{13}+\frac{\delta Z_{31}}{2}), \\
\delta \Gamma_{H_{2}tt}^{L}
&=
g_{H_{1}tt}^{L}\qty(\delta \theta_{12}+\frac{\delta Z_{12}}{2})\qc
\delta \Gamma_{H_{2}tt}^{R}
=
g_{H_{1}tt}^{R}\qty(\delta \theta_{12}+\frac{\delta Z_{12}}{2}), \\
\delta \Gamma_{H_{3}tt}^{L}
&=
g_{H_{1}tt}^{L}\qty(\delta \theta_{13}+\frac{\delta Z_{13}}{2})\qc
\delta \Gamma_{H_{3}tt}^{R}
=
g_{H_{1}tt}^{R}\qty(\delta \theta_{13}+\frac{\delta Z_{13}}{2}),
\end{align}
where $g_{H_{i}tt}^{L/R}$ are the top-quark Yukawa couplings at the tree-level. In our scenario, they are obtained as
\begin{align}
g_{H_{1}tt}^{L} = g_{H_{1}tt}^{R} = -\frac{m_{t}}{v}\qc
g_{H_{2}tt}^{L} = -ig_{H_{3}tt}^{L} = -\frac{\rho_{tt}^{*}}{\sqrt{2}}\qc
g_{H_{2}tt}^{R} = -ig_{H_{3}tt}^{R} = -\frac{\rho_{tt}}{\sqrt{2}}.
\end{align}
Following ref.~\cite{Kanemura:2004mg}, we determine the mixing counter terms $\delta\theta_{ij}$ by using the relations with $\delta Z_{ij}$.
In our scenario, they are obtained as
\begin{align}
    \delta \theta_{ij}
    &=
    -\frac{1}{2}\qty(\frac{\delta Z_{ij}}{2}-\frac{\delta Z_{ji}}{2}).
    \label{eq: del_thetaij_KOSY_minimal}
\end{align}
From eqs.~\eqref{eq: delZ_ij} and \eqref{eq: del_thetaij_KOSY_minimal}, we obtain
\begin{align}
-\delta \theta_{12}+\frac{\delta Z_{21}}{2}
&=
\delta \theta_{12}+\frac{\delta Z_{12}}{2}
\notag \\
&=
\frac{3\lambda_{3}\lambda_{7}^{R}v^{2}}{2(m_{\Phi}^{2}-m_{H_{1}}^{2})}\qty(B_{0}[m_{H_{1}}^{2}; m_{\Phi}^{2}, m_{\Phi}^{2}]-B_{0}[m_{\Phi}^{2}; m_{\Phi}^{2}, m_{\Phi}^{2}]), \\
-\delta \theta_{13}+\frac{\delta Z_{31}}{2}
&=
\delta \theta_{13}+\frac{\delta Z_{13}}{2}
\notag \\
&=
-\frac{3\lambda_{3}\lambda_{7}^{I}v^{2}}{2(m_{\Phi}^{2}-m_{H_{1}}^{2})}\qty(B_{0}[m_{H_{1}}^{2}; m_{\Phi}^{2}, m_{\Phi}^{2}]-B_{0}[m_{\Phi}^{2}; m_{\Phi}^{2}, m_{\Phi}^{2}]).
\end{align}
Then, from figure~\ref{fig:CEDMdiagram3}, we obtain
\begin{align}
    \tilde{d}_t^{(2), \mathrm{OS}}
    &=
    \tilde{d}_t^{(2)}+\Delta \tilde{d}_t^{(2), \mathrm{OS}},
    \notag \\
    \Delta \tilde{d}_t^{(2), \mathrm{OS}}
    &=
    \frac{\mathrm{Im}[\lambda_7\rho_{tt}]}{\sqrt{2}} \frac{3 \lambda_3 v}{(16 \pi^2)^2} \frac{2m_t^2}{m_{\Phi}^2 - m_{H_1}^2}
    \Big(C_{11}[\Phi, t, t] - C_{11}[H_1, t, t] \Big)
    \notag \\ &\quad\times
    \qty(
    2B_{0}[0; m_{\Phi}^{2}, m_{\Phi}^{2}]
    -B_{0}[m_{H_{1}}^{2}; m_{\Phi}^{2}, m_{\Phi}^{2}]
    -B_{0}[m_{\Phi}^{2}; m_{\Phi}^{2}, m_{\Phi}^{2}]
    ),
\end{align}
where $\tilde{d}_t^{(2)}$ is given in eq.~\eqref{eq:topCEDM2} while it is evaluated with the OS quantities not the EP quantities. 

To understand the origin of $\Delta \tilde{d}_t^{(2), \mathrm{OS}}$, we discuss the relation of the mixing angles $\theta_{ij}$ in the EP and OS schemes.
From eq.~\eqref{eq:relationRM}, we obtain the mixing counter terms in the EP scheme as
\begin{align}
\delta \theta_{12}^{\mathrm{EP}} 
&= -\frac{\delta \Pi_{12}}{m_{\Phi}^{2}-m_{H_{1}}^{2}}
= \frac{\Pi_{12}^{\lambda_{7}}(0)}{m_{\Phi}^{2}-m_{H_{1}}^{2}}+\dots~,
\notag \\
\delta \theta_{13}^{\mathrm{EP}} 
&= -\frac{\delta \Pi_{13}}{m_{\Phi}^{2}-m_{H_{1}}^{2}}
= \frac{\Pi_{13}^{\lambda_{7}}(0)}{m_{\Phi}^{2}-m_{H_{1}}^{2}}+\dots~,
\label{eq:EPdeltatheta}
\end{align}
where $\delta \Pi_{12}$ and $\delta \Pi_{13}$ are given in eq.~\eqref{eq:Piij_CT}, and we have only shown the EDM relevant parts.
Then, from eqs.~\eqref{eq: delZ_ij}, \eqref{eq: del_thetaij_KOSY_minimal}, \eqref{eq:EPdeltatheta}, and the relation
\begin{align}
    \theta_{ij}^{\mathrm{OS}} = \theta_{ij}^{\mathrm{EP}} + \delta \theta_{ij}^{\mathrm{EP}} - \delta \theta_{ij}^{\mathrm{OS}}, 
\end{align}
we obtain
\begin{align}
    \theta_{12}^{\mathrm{OS}}
    &=
    \theta_{12}
    +\frac{2\Pi_{12}^{\lambda_{7}}(0)-\Pi_{12}^{\lambda_{7}}(m_{H_{1}}^{2})-\Pi_{12}^{\lambda_{7}}(m_{\Phi}^{2})}{2(m_{\Phi}^{2}-m_{H_{1}}^{2})}+\dots~ \eval{}_{\mathrm{EP}},
    \notag \\
    \theta_{13}^{\mathrm{OS}}
    &=
    \theta_{13}
    +\frac{2\Pi_{13}^{\lambda_{7}}(0)-\Pi_{13}^{\lambda_{7}}(m_{H_{1}}^{2})-\Pi_{13}^{\lambda_{7}}(m_{\Phi}^{2})}{2(m_{\Phi}^{2}-m_{H_{1}}^{2})}+ \dots~\eval{}_{\mathrm{EP}},
    \label{eq:thetaijOStoEP}
\end{align}
where everything in the right-hand side is the EP quantities.
In our scenario, we take $\theta_{ij}=0$ and there is no one-loop contribution in the EP scheme (see eq.~\eqref{eq:topCEDM1}).
On the other hand, in the OS scheme, we have the one-loop contributions.
\begin{align}
\tilde{d}_{q}^{(1), \mathrm{OS}}
&=
-\frac{4m_{t}^{2}}{16\pi^{2}}\frac{1}{\sqrt{2}v}\qty(
\Im[\rho_{tt}^*]\theta_{12}^{\mathrm{OS}}+\Im[i\rho_{tt}^*]\theta_{13}^{\mathrm{OS}}
)\Big(C_{11}[\Phi,t,t]-C_{11}[H_{1},t,t]\Big),
\label{eq: topCEDM1_OS}
\end{align}
where we have assumed that $\theta_{ij}^{\mathrm{OS}}$ are small in eq.~\eqref{eq:topCEDM1}.
By substituting eq.~\eqref{eq:thetaijOStoEP} into eq.~\eqref{eq: topCEDM1_OS}, we obtain
\begin{align}
    \tilde{d}_t^{(1),\mathrm{OS}}+ \tilde{d}_t^{(2),\mathrm{OS}}
    =
    \eval{\tilde{d}_t^{(2)}+\mathcal{O}(\hbar^{3})}_{\mathrm{EP}}.
\end{align}
Thus, the difference in the two-loop expressions corresponds to the scheme difference of the mixing angles, and we again obtain the same prediction up to the two-loop order.

\bibliographystyle{utphys28mod}
\bibliography{references}

\end{document}